\documentclass[preprint]{aastex}

\newcommand{\subsun}{\mbox{$_{\odot}$}}
\newcommand{\etal}{{\it et al.\/}}
\begin{document}

\title{The Ages and Abundances of a Sample of Globular Clusters in 
M49 (NGC 4472)\altaffilmark{1}}

\author{Judith G. Cohen\altaffilmark{2}, J.~P. Blakeslee\altaffilmark{3}
\& P. C\^ot\'e\altaffilmark{4}}

\altaffiltext{1}{Based on observations obtained at the
W.M. Keck Observatory, which is operated jointly by the California 
Institute of Technology, the University of California, and the
National Aeronautics and Space Administration.}
\altaffiltext{2}{Palomar Observatory, Mail Stop 105-24,
California Institute of Technology, Pasadena, Ca., 91125}
\altaffiltext{3}{Department of Physics and Astronomy, Center for
Astrophysical Sciences, Johns Hopkins University, 3400 North Charles Street,
Baltimore, MD 21218}
\altaffiltext{4}{Department of Physics and Astronomy, Rutgers University,
136 Frelinghuysen Road, Piscataway, N.J. 08854-8019}

\begin{abstract}
We present a study of the metallicity and age of the globular clusters
system of M49 (NGC 4472), the most luminous galaxy in the Virgo cluster. 
We measure Lick indices for 47 GCs in M49 from LRIS/Keck spectra and
establish their metallicity parameters qualitatively in comparison to the
Galactic GCs and to published data for 150 M87 GCs.
We then compare our measurements with the predictions of models for the integrated
light of old single burst stellar systems by \cite{worthey94b}
and by \cite{thomas02}.
We find
that the metallicity of the M49 GC system spans the range from 
[Fe/H]$_Z = -2.0$ to +0.4 dex.  We show that the
metallicity and age parameters for these
two GC systems are basically identical, except that
the M49 GCs reach slightly higher metallicities than 
do those of M87.  We find that the  GCs of both of these 
giant elliptical galaxies
are $\alpha$-enhanced by a factor of about two above the Solar value, as is
also true of the Galactic GCs.  Thus
the most metal rich GCs in M49 reach [Z/H] $\sim$ +0.8 dex, comparable
to that of M49 itself.  While adoption
of the $\alpha$-enhanced models of \cite{thomas02} has eliminated most
of the previous discrepancies with observations, 
the most metal rich M49 GCs have NaD lines which are still considerably
stronger than those predicted by any model, and there are still issues
involving the metallicity scale of these models.  We find 
that in the mean, the M49
GCs are at least 10 Gyr old.  However, the grids of models we used
differ in how they treat the horizontal branch, and this perceptibly
affects the predicted H$\beta$ index.  Hence our current 
incomplete understanding of
the role of the horizontal branch limits our ability to derive a more detailed
distribution for the ages of the GCs in M49 and in M87.

\end{abstract}


\section{Introduction}

M49 (NGC 4472) is the most luminous galaxy in the Virgo cluster, 
with $M_V = -22.6$, and is more luminous than any closer galaxy.
It is, however, not located at or near the dynamical center
of the cluster.  Instead, M87 resides at the center of this
deep potential well, surrounded by a huge
amount of X-ray emitting gas, while M49 is 
at the center of the modest Virgo B subcluster
more than 4.5$^{\circ}$ away.  The globular
cluster (henceforth GC) system of M49 is rich, with S/N = 3.6$\pm0.6$ 
($\sim5900$ GCs in total)\citep{rhode01},
but that of M87 is even richer (S/N = 14$\pm4$, 
with $\sim$13,450 GCs in total) \citep{mclaughlin94}.  
The enormous GC system of M87 made it a tempting
target, and JGC and collaborators carried out an exhaustive study
of the kinematics of the GC system of M87 \citep{cohen97,cohen2000}
and of the ages and abundances of a large sample
of M87 GCs \citep{cohen98}.  We then felt
we needed to investigate the GC system of M49 to understand which attributes
of the GC system of M87 might be unique to its special location, and which
generic to luminous giant ellipticals.
Our analysis of the kinematics of the GC system of M49
and inferences concerning the distribution of mass and dark matter in M49
and the isotropy of the GC orbits is given in \cite{cote03}, while the most
recent radial velocity tabulation and kinematic analysis for the M87
GC system can be found in \cite{hanes01} and \cite{cote01}.  Here we
utilize the best of our spectra to
explore the distribution in age and in metallicity 
as well as the abundance ratios of the GC system of M49.

The first attempts to determine the metallicity of the Virgo GC system
were based on wide field photometry.  This began with the work of 
\cite{strom81} for M87 and that of \cite{cohen88} for M49,
but we jump to the ground based wide field
Washington photometry of \cite{geisler96} and to the
HST based analysis of \cite{whitmore95}.  In these recent studies,  
the key new result was the detection of strong bimodality in
the color distribution of the GCs.  Although no separation of age
and metallicity was possible, the 
observed color distribution of the GCs was assumed to reflect the
intrinsic metallicity distribution. Such a bimodal metallicity
distribution might arise through GC formation in major mergers, through
the hierarchical assembly of the galaxy through numerous
dissipationless mergers, or through multiple, internally-driven bursts
of star and cluster formation (for a review of the various formation
models, see Carney and Harris 2001).

\cite{mould90} obtained the first spectra of  GCs in M49
and from their sample of 26 GCs
suggested the mean metallicity of the M49 GC system was 1/6 Solar.
\cite{sharples98} presented spectra and a kinematic
analysis for an additional 47 GCs in M49.  Spectra of another 87 GCs
in M49 were obtained by \cite{zepf00}, who presented a kinematic analysis.
\cite{beasley00} carried out an analysis of the available spectra for
131 GCs in M49.  However,
their spectra have low signal-to-noise ratios, and they
could not derive any relevant parameters for individual M49 GCs.  Instead,
they summed the spectra into four bins in the
$C-T_1$ color of the Washington photometry of \cite{geisler96},
assuming this correlated with metallicity, to derive mean abundances
and ages.

The advantage of LRIS \citep{oke95}, which is a multi-slit instrument 
utilizing  
the large collecting area of the 10-m Keck Telescope, in this situation
is crucial.  We are able to derive credible parameters for individual
GCs in Virgo cluster galaxies.  
Since the work of \cite{cohen98}, the sophistication of the available
models of the integrated light of old stellar systems has increased,
and we  explore here the issue of possible $\alpha$-enhancement
(non-Solar ratios of Mg/Fe or Na/Mg) in both M87 and in M49 as well
as the metallicity  
(i.e. [Fe/H]\footnote{The 
standard nomenclature is adopted; [X/H] = log$_{10}$[N(X)/N(H)] $-$ log$_{10}$[N(X)/N(H)]\subsun, and similarly
for [X/Fe].}) and age of the GCs in these two galaxies.

\section{Selection of Candidates and Observations}

The photometric study of the GC system of M49 (NGC~4472) 
with the intermediate band
Washington filter system by \cite{geisler96} was used to define
the sample of candidate GCs, which are identified using their
numbering system.  Coordinates for the candidates
were determined as described in \cite{cote03}.  A color cut
of $1.0 \le (C-T_1) \le 2.25$ mag, similar to that used by
\cite{geisler96}, was imposed to avoid contamination
by foreground Galactic stars or background galaxies.  In practice,
essentially no genuine GC candidates are found outside this range
\citep{lee93,ostrov98}.
For the present work, we preferentially selected
those spectroscopically confirmed GCs in M49 (either from 
our own radial velocity catalog as it existed in early 2000 or
from the literature)
which were bright, with $T_1 < 20.5$, where $T_1$ is very close to 
$R$ \citep{geisler90}.
The positions of these bright GCs were then used to design two LRIS
slitmasks. Additional bright candidate GCs from the database of 
\cite{geisler96} were added to maximize the number of slits
per mask.

The two slitmasks were observed with LRIS \citep{oke95}
at the Keck Telescope on the night of May 1, 2000. The weather
conditions were good, with seeing 0.7 to 0.8 arcsec.  A
600 g/mm grating blazed at 5000 \AA\ (spectrum centered near
5400~\AA, and covering roughly the range from H$\beta$
to H$\alpha$ for a slitlet near the centerline of the slitmask)
gave a spectral resolution of 1.2~\AA/pixel, or 5.8~\AA\
for the 1.0 arcsec wide slitlets.  Three 3000 sec exposures were
obtained for the first slitmask and two for the second.
The spectra from this run, which are the only ones utilized in the present
paper, are the best of the 196 new spectra for M49 GCs discussed
in \cite{cote03}.

The spectra were reduced using Figaro \citep{shortridge93} scripts
developed for analysis of LRIS multi-slit spectra
as described in our work on the GC system of M87 \citep{cohen97}.
One improvement from our earlier work was in the wavelength calibration
of the spectra.  As before, 
we used the a third order polynomial fit to the
night sky emission lines in the spectrum from each slitlet to calibrate
the wavelength scale.  A problem encountered in our previous
work on M87 was the lack of any detectable night sky lines
blueward of the 5199~\AA\ [N~I] line.  (Keck spectra
do not show Hg or other emission features from city lights in
one hour exposures with LRIS.)
Since we wished to examine the region around H$\beta$, we were forced
to extrapolate the wavelength fit uncomfortably far; such a procedure
sometimes leads to runaways in the fit in the unconstrained region at 
the beginning or end of the spectrum. Thus in the present work
a search was made for additional night sky emission lines in these spectra  
blueward of 5199~\AA.  A very weak line
was found at 4968.8~\AA, whose reality was checked on several
spectra of other objects.  With this additional line, arising from
a transition in O~I, the
wavelength scale is now more robust at H$\beta$.

The radial velocities of the M49 GC candidates were determined
as described in \cite{cohen97}.
Typical  $1\sigma$ errors in the wavelength fit are less than 0.1~\AA;
these are a minor contributor to the assigned $v_r$ errors, which are
based on the signal-to-noise ratio (SNR) of the spectra.  Comparison
of the $v_r$ measured from various
features suggests that these assigned errors 
are quite conservative.  The radial velocities from
these observations
are included as part of our larger database on GCs in M49
presented in \cite{cote03} and are tabulated there.

The resulting 1D spectra, summed along the slit and over the
available exposures, cannot be fluxed effectively due to the
variable slit losses  inherent in use of a multislit mask with
fairly narrow slits.

\subsection{Foreground and background objects}

Geisler 1982 appeared extended on the slitmask alignment 
setup images and its spectrum is that of a galaxy.
The redshifts of four additional galaxies which
turned up serendipitously in the slitlets, as well as that of Geisler 1982,
are given in Table~\ref{table_gals}.  The $R$ mags given in the table
are from our photometry carried out on the slitmask alignment images
taken with LRIS.  Only one Galactic star,
Geisler 2860, was found among the objects observed.
After elimination of foreground and background objects, we have a sample
of 47 GCs in M49 with LRIS spectra.

\section{Measured Indices}

We have measured the standard Lick indices \citep{faber85,burstein84}
as well as the H$\alpha$ index 
defined in our earlier work \citep{cohen98} on the M87 GC system.
The codes with which this
was accomplished are the same Figaro \citep{shortridge93} routines
used in \cite{cohen98}.  These take the standard bandpasses, shift
them in accordance with the radial velocity of each object, then
calculate the indices.

The results of this automatic procedure were examined, and 
discrepant and/or problem
cases were checked individually.  The NaD indices in these faint objects
are particularly difficult as the corresponding emission 
lines from the night sky must be removed.  Some of the spectra
were rereduced manually in an effort to improve the night sky subtraction. 
H$\alpha$ is beyond the red end of the spectra for a few of the
M49 GC candidates; for several others, this index had to be measured
by hand when the red continuum bandpass for this feature
was only partially included.

The resulting set of indices are listed in Table~\ref{table_indices}.
Because the spectra are not fluxed and are not matched to the spectral
response used to define the Lick system
as no Lick standard stars were observed, those indices which span
wide ranges in wavelength may have systematic offsets from the nominal
Lick system.  The narrower feature indices which we will use, 
such as H$\beta$, Mgb, Fe5270
and NaD,  are not significantly affected by this concern.

The last column of this table gives the SNR per 
spectral pixel in the final spectrum,
measured over the interval 5950 to 6250~\AA.  Within this
wavelength interval the observed spectra are approximately 
constant in DN/pixel and there are few strong spectral features.
The signal level in the continuum at H$\beta$ is 
60 to 75\% of that at 6100~\AA.  The errors for the indices were
computed from these measurements of the SNR in the continuum.
 This too is a departure from our earlier work, where the SNR was 
{\it{calculated}} from the signal level in the spectrum and in  
the background.  The sky and the spatially variable light
from the halo of M49 contribute to the background.
Furthermore slit losses depend
on the accuracy of the coordinates and the alignment of the slitmask.
Hence the SNR may not be a smoothly increasing
function of the brightness of the GC candidate.

\section{Qualitative Analysis}

We begin with a qualitative examination of our data, considering
the correlations among the indices measured for the M49 GC
sample,  comparing  with those found in our
earlier study of the  M87 GCs \citep{cohen98} and also
with those of the Galactic GC system.

\subsection{Correlations of Indices}

\begin{figure}
\epsscale{0.9}
\plotone{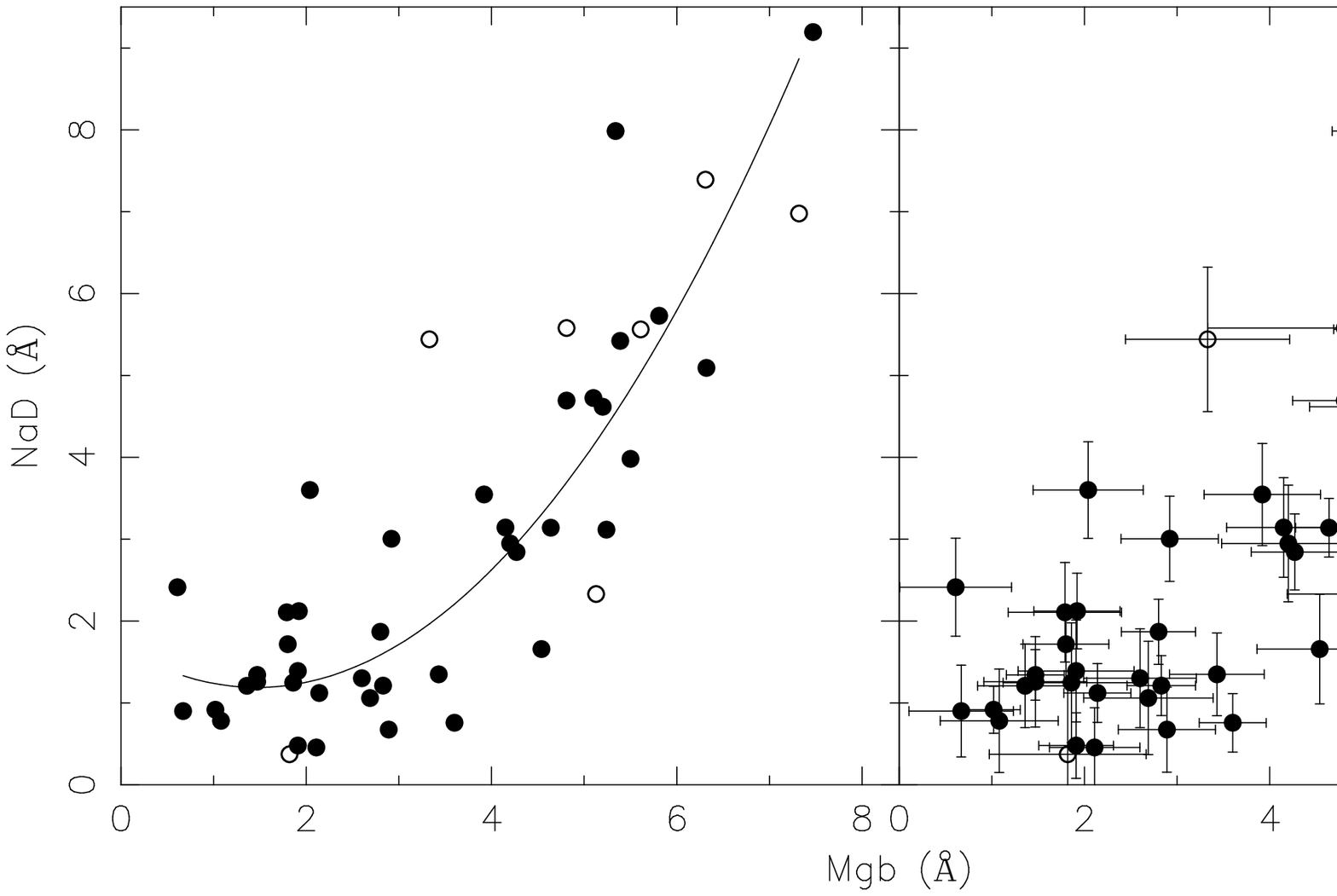}
\caption[]{The NaD index is shown as a function of the Mgb index for the 47
GCs in M49.  Objects with $SNR < 10$ are indicated by open circles.
In the left panel, the solid curve is the weighted second order fit
to all the M49 GCs.
In the right panel, 1$\sigma$ error bars are shown for each GC.
\label{figure_mgna}}
\end{figure}

We display the correlation of NaD with Mgb in Figure~\ref{figure_mgna}.
The errors in each index, computed as described above, are indicated
in the right panel of this figure.
The expected correlation of increasing NaD with increasing Mgb,
presumably arising primarily from increasing metallicity, is seen.
The coefficients of the weighted second order polynomial fit
to this relationship, shown in the right panel of this figure,
are given in Table~\ref{table_coef}. This fit is shown over the
range which includes 95\% of the GC candidates, excluding 1 GC
at each end\footnote{A larger sample, such as that of \cite{cohen98} in M87,
would make this measure more robust and less affected by isolated
outliers or bad measurements.}.  The limits
of this range for each index are given in Table~\ref{table_range}.    
The dispersion about this fit, assuming no errors in Mgb, is 
1.16~\AA,  
consistent with the uncertainties in the NaD indices of the
M49 GCs and the problems of subtraction of the night sky emission
that affect this index.  

\begin{figure}
\epsscale{0.9}
\plotone{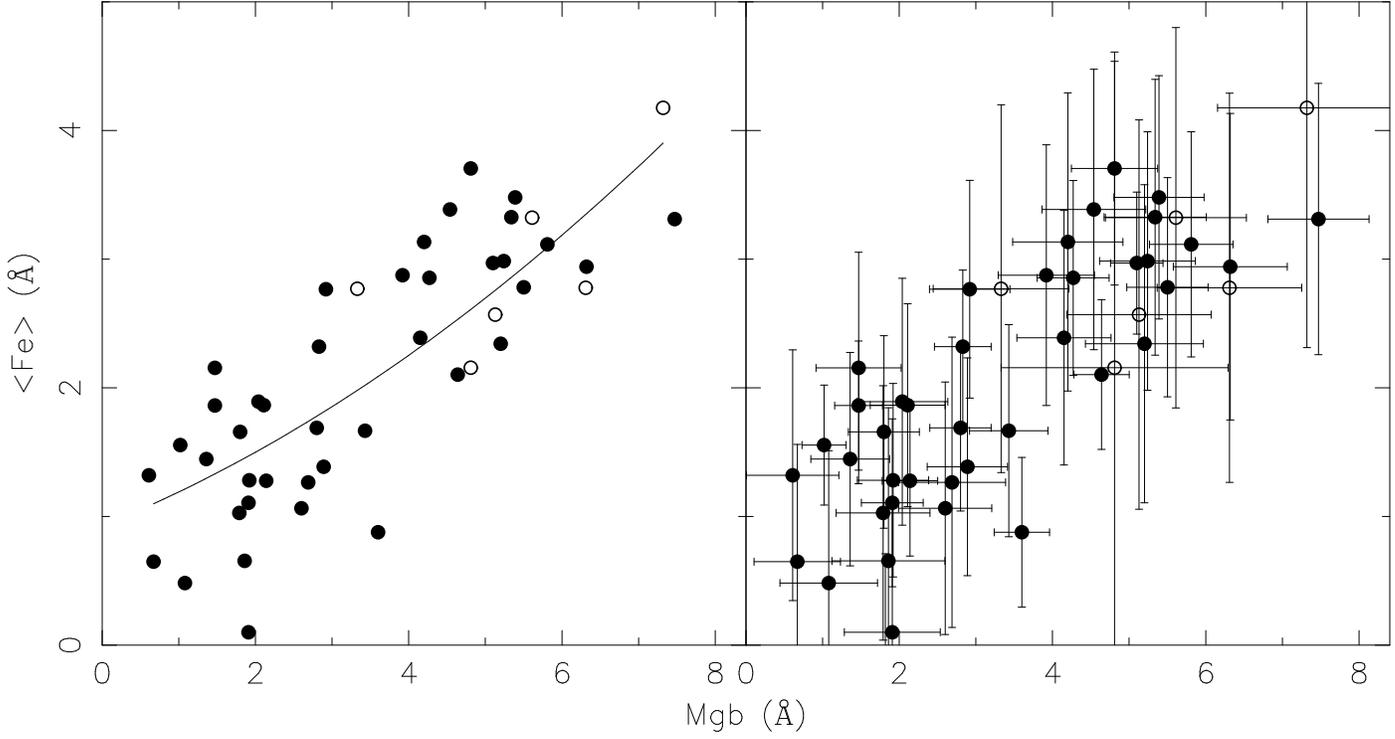}
\caption[]{The average of the Fe5270 and Fe5335 index
is shown as a function of the Mgb index for the 47
GCs in M49.  Objects with $SNR < 10$ are indicated by open circles.
In the left panel, the solid curve is the weighted second order fit
to all the M49 GCs.
In the right panel, 1$\sigma$ error bars are shown.
\label{figure_mgfe}}
\end{figure}

We display the correlation of the mean of the 
Fe5270 and the Fe5335  indices (denoted $<$Fe$>$) with 
the Mgb index in Figure~\ref{figure_mgfe}. 
The expected increase of $<$Fe$>$ with increasing Mgb index,
presumably arising primarily from increasing metallicity, is seen in this figure.
The coefficients of the weighted second order polynomial fit
to this relationship are given in Table~\ref{table_coef} and the
95\% range is again given in Table~\ref{table_range}.    The
dispersion about this fit, assuming no errors in Mgb, is 
only 0.6~\AA, consistent with the uncertainties in the 
Fe5270 and Fe5335 indices of the M49 GCs.

\begin{figure}
\epsscale{0.9}
\plotone{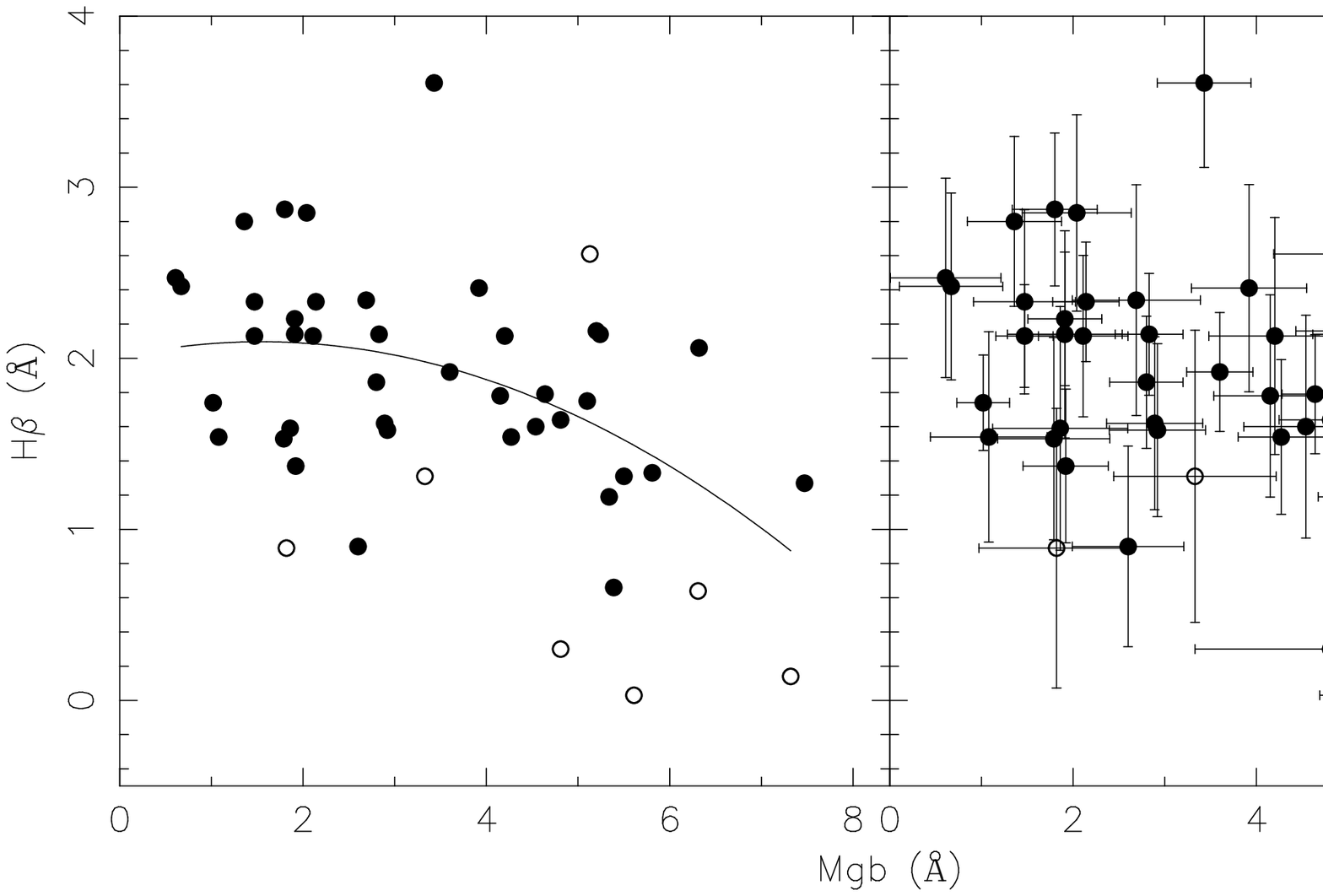}
\caption[]{The H$\beta$ index
is shown as a function of the Mgb index for the 47
GCs in M49.  Objects with $SNR < 10$ are indicated by open circles.
In the left panel, the solid curve is the weighted second order fit
to all the M49 GCs.
In the right panel, 1$\sigma$ error bars are shown.
\label{figure_mghb}}
\end{figure}

The correlation of H$\beta$ with Mgb is shown in Figure~\ref{figure_mghb}.
H$\beta$ is the bluest index measured, and has the lowest signal
level in the continuum of any of the measured indices.  Since H$\beta$ in
high metallicity old GCs is expected to be weak, 
we anticipate and we see substantial
scatter in this figure.  However, these two indices are clearly 
anti-correlated,
as is expected from models of old single burst stellar populations.
The coefficients of the weighted second order polynomial fit
to this relationship are given in Table~\ref{table_coef} and the
range is again given in Table~\ref{table_range}.  The
dispersion about this fit, assuming no errors in Mgb, is 
only 0.6~\AA, consistent with the uncertainties in the H$\beta$ indices of the
M49 GCs.

\begin{figure}
\epsscale{0.9}
\plotone{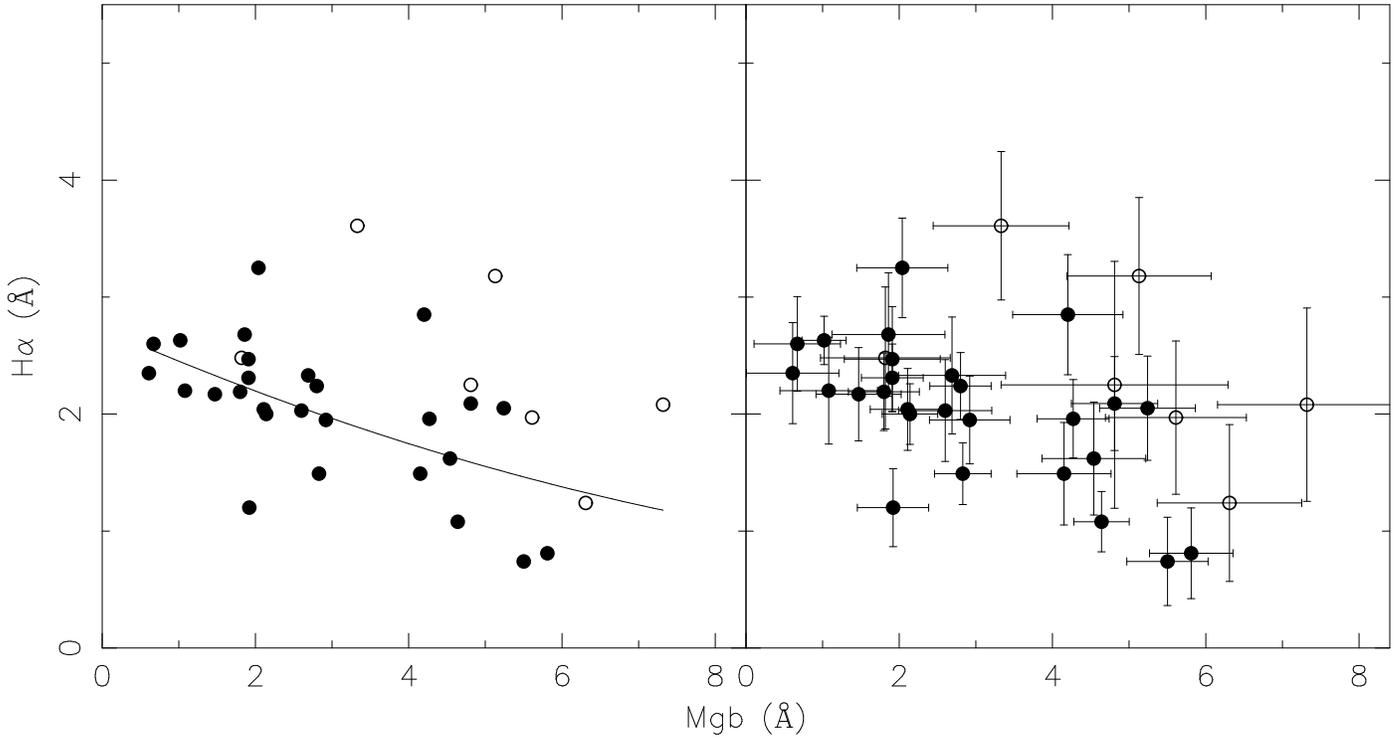}
\caption[]{The H$\alpha$ index, detected in 34 of the M49 GCs,
is shown as a function of the Mgb index.  
Objects with $SNR < 10$ are indicated by open circles.
In the left panel, the solid curve is the weighted second order fit
to all the M49 GCs.
In the right panel, 1$\sigma$ error bars are shown.
\label{figure_hamg}}
\end{figure}

The correlation of H$\alpha$ with Mgb is shown in 
Figure~\ref{figure_hamg} for the 34 GCs in our sample with
detected H$\alpha$ lines.  The expected correlation 
between the two indices is seen.
The coefficients of the weighted second order polynomial fit
to this relationship are given in Table~\ref{table_coef} and the
95\% range is again given in Table~\ref{table_range}.  The
dispersion about this fit, assuming no errors in Mgb, is 
only 0.6~\AA, consistent with the uncertainties in the H$\alpha$ indices of the
M49 GCs.

\begin{figure}
\epsscale{0.9}
\plotone{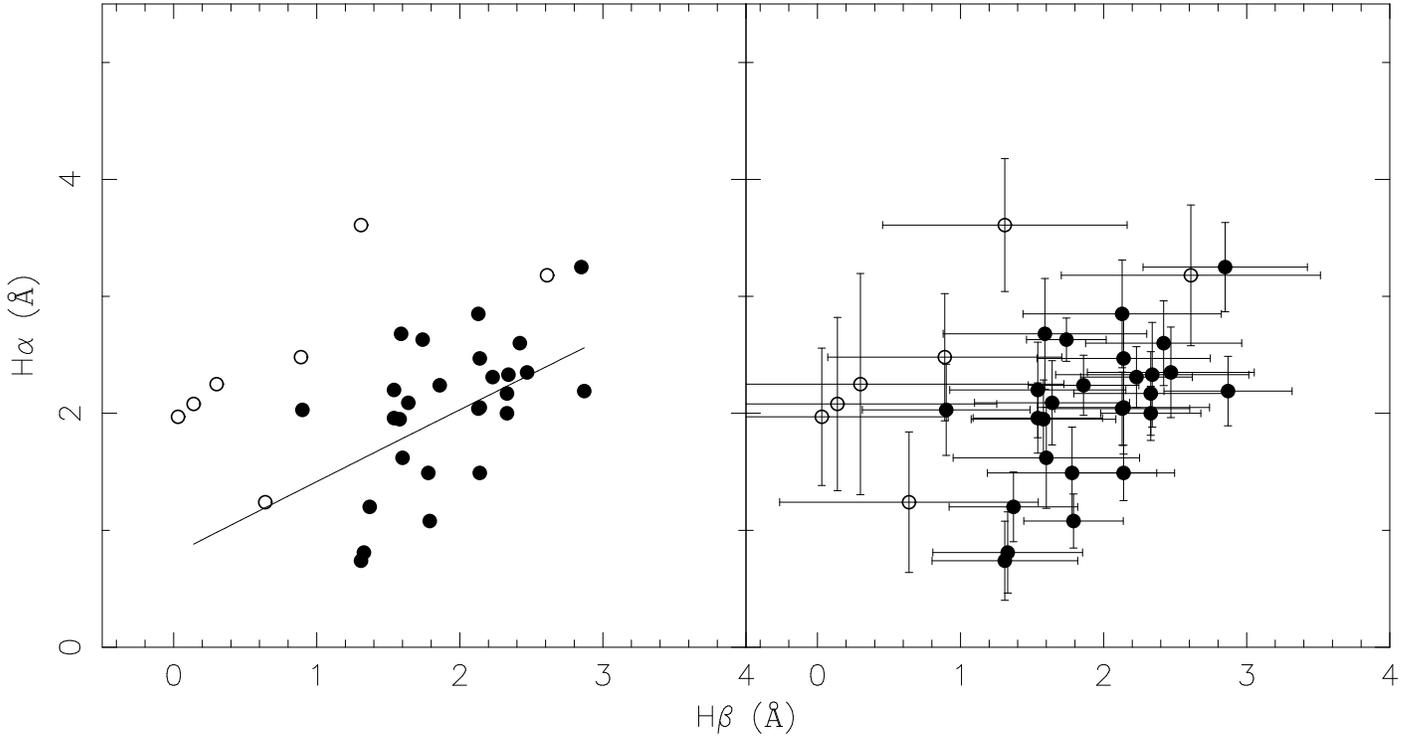}
\caption[]{The H$\alpha$ index
is shown as a function of the H$\beta$ index for the 
GCs in our sample in M49 where H$\alpha$ was detected (34 GCs).  
Objects with $SNR < 10$ are indicated by open circles.
In the left panel, the solid curve is the weighted second order fit
to the high SNR data only.
In the right panel, 1$\sigma$ error bars are shown.
\label{figure_hahb}}
\end{figure}

The correlation of the H$\alpha$ indices of the M49 GCs with their
H$\beta$ indices is shown
in Figure~\ref{figure_hahb}.  The expected increase of 
H$\alpha$ with increasing H$\beta$ is seen.
The coefficients of the weighted second order polynomial fit
to this relationship are given in Table~\ref{table_coef} and the
range is again given in Table~\ref{table_range}.    The
dispersion about this fit, assuming no errors in H$\beta$, is 
0.45~\AA, consistent with the uncertainties in the H$\alpha$ indices of the
M49 GCs.

\subsection{Comparison with M87}

\begin{figure}
\epsscale{0.9}
\plotone{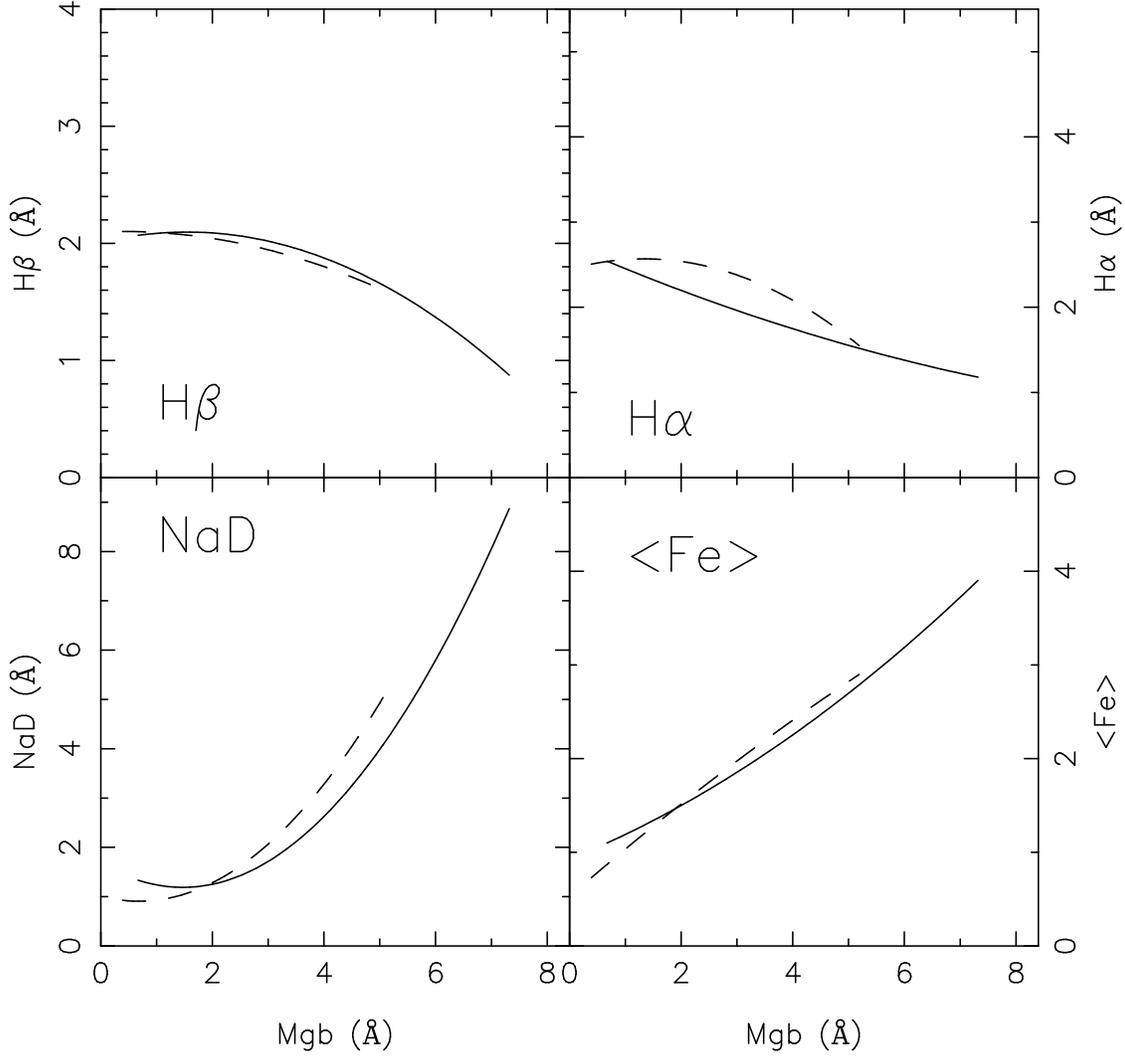}
\caption[]{The four panels display the
second order fits to the index-index plots 
NaD, $<$Fe$>$,
H$\beta$ and H$\alpha$ as a function of Mgb shown in
Figures 1--4 for the M49 GCs (solid curves) and for those of M87 
from \cite{cohen98}(dashed curves).
The curves extend over the 95\% range in each case.
\label{figure_m87comp}}
\end{figure}

The general correlations among the Lick indices
that we see in the GC system of M49 were also found
by \cite{cohen98} to hold in the M87 GC system.
Figure~\ref{figure_m87comp} shows the best fit second
order curves to the index-index plots NaD, $<$Fe$>$,
H$\beta$ and H$\alpha$ as a function of Mgb from 
Figures 1 to 4 for M49.  The fits to the indices of the 150 GCs in
M87 from \cite{cohen98} are also shown.  
These fits are displayed over the range which includes 95\% of
the GCs in our sample in each system.
These ranges for the M87 sample with suitable LRIS spectra
are given in Table~\ref{table_range}.

The radial distribution of the spectroscopic sample in M49
and in M87 is compared in Table~\ref{table_radial}.  The
final column gives the median galactocentric radius $R_{GC}$
in units of the galaxy scale length. 
The minimum $R_{GC}$ 
and $R_{GC}$  of the first quartile are
comparable for the two spectroscopic GC samples.  However, the M87 GC
spectroscopic sample extends to larger $R_{GC}$.  Since the radial
gradient in GC properties is small in the outer region of
M87 \citep*[e.g.][]{cohen98}, this difference in sample spatial 
distribution should
not cause a significant bias in the relative abundance
distributions of the GC systems in these two galaxies

We see in Figure~\ref{figure_m87comp} that
the fits to the index-index correlations for the two Virgo
gEs are essentially  identical.  However, the upper range limits  are larger 
for each of the major metallicity indices
Mgb, NaD and $<$Fe$>$ for M49 than for M87.  Following
upon our previous work in M87, we thus
assert that to first order, both systems are composed
of predominantly old clusters spanning a wide range
of metallicity, with the range of metallicity 
extending to a somewhat higher value in M49 than in M87.

\subsection{Comparison with Galactic Globular Clusters}

We next compare the behavior of the index-index plots 
for the M49 GCs with those of the Galactic globular clusters.
The indices for the Galactic globular clusters are taken
from \cite{cohen98}, \cite{puzia02}, \cite{covino95} 
or \cite{burstein84} (in 
decreasing order of preference for clusters with multiple
observations).  This gives a sample of 35 Galactic GCs 
observed, spanning the full range of metallicity of
the Galactic GC system.  In particular, with the
work of \cite{cohen98} and subsequently of \cite{puzia02},
the most metal rich Galactic GCs and those in
the bulge are well sampled, while \cite{covino95} concentrated
on the metal poor end.

\begin{figure}
\epsscale{0.9}
\plotone{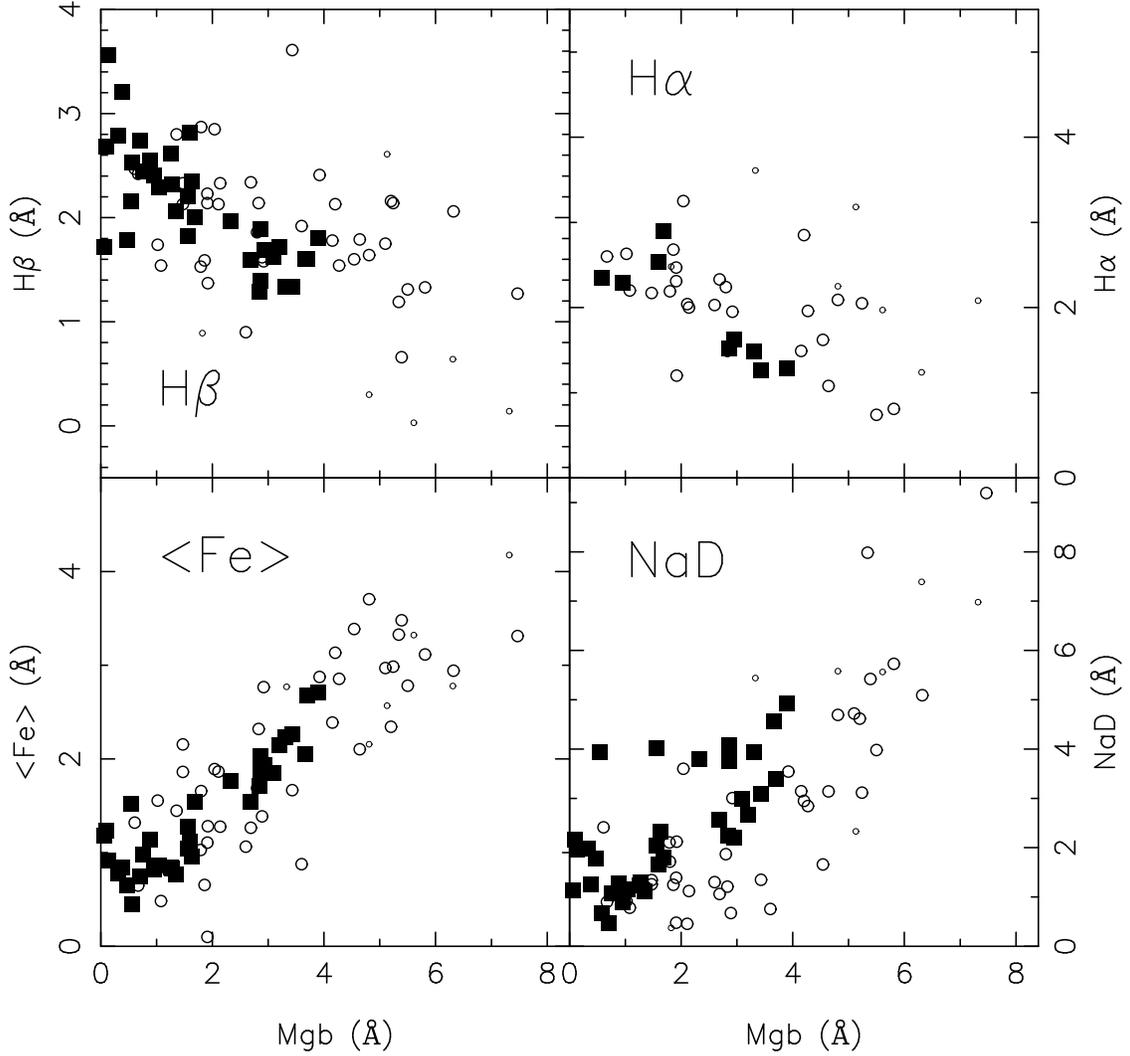}
\caption[]{The four panels display the NaD, $<$Fe$>$, H$\beta$
and H$\alpha$ indices 
as a function of the Mgb index for the 47
GCs in M49 (large open circles), only 34 of which have detections at
H$\alpha$.  Small open circles denote the
objects with $SNR < 10$.  The Galactic globular clusters
with data from \cite{cohen98}, \cite{puzia02}, \cite{covino95} or \cite{burstein84}
are shown as filled squares.
\label{figure_quadgalglobs}}
\end{figure}

Figure~\ref{figure_quadgalglobs} shows the relationships  
for the index pairs Mgb-NaD, Mgb-$<$Fe$>$, Mgb-H$\beta$
and H$\alpha$-H$\beta$ for the M49 and for the galactic GC samples.
In each case, the behavior is
similar, but the M49 GCs reach higher values of the metallicity
sensitive indices than do the most metal rich Galactic GCs.
The comparison of the anti-correlation between
H$\beta$ and Mgb suggests that the age of the M49 GCs
is comparable to that of the Galactic GCs, which we
take to be (almost all) uniformly old, with an age of
12 Gyr \citep*[see, e.g.][]{rosenberg99}\footnote{We ignore
the GCs associated with the accretion of the Sgr dwarf galaxy
\citep{ibata97}, which may be $\sim$3 Gyr younger \citep{fusipecci95}.}.

Furthermore, although there is extremely good agreement in 
the mean relationship between Mgb and $<$Fe$>$ indices
between the M49 and the Galactic GCs,
there is a systematic vertical offset upward
(with some scatter) in the panel of
Figure~\ref{figure_quadgalglobs} displaying Mgb vs. NaD.
We suggest that this arises due to  NaD 
absorption from the ISM in the disk
of our Galaxy; this absorption  is presumably absent in
M49, which has essentially no gas and no disk.
This effect will appear large both for GCs with high reddening
and also for the more metal poor clusters,
where even a small contribution from the ISM can perturb the weak
NaD lines in the integrated light of the cluster.
A similar offset is seen for the M87 GCs \citep*[see figure 9 of][]{cohen98}.

\subsection{``New'' versus ``Old'' Lick Indices}

To maintain consistency with our earlier work on the M87 GC system
\citep{cohen98},
we have used the standard Lick indices \citep{faber85,burstein84}
together with an H$\alpha$ index
defined in \cite{cohen98}.  We can then use with confidence
the transformation previously determined by \cite{cohen98}
between [Fe/H]$_W$, the metallicity inferred directly from the
\cite{worthey94b} grid of models, and [Fe/H]$_Z$, that
metallicity on the well calibrated and commonly used \cite{zinn84} scale.
The exact definition of the wavelength bandpasses for
the Lick indices has evolved somewhat with time.  \cite{worthey94a}
presented updated definitions of these indices which are slightly
wider to better accommodate the high velocity dispersions 
found in massive galaxies.  For a few of these indices,
\cite{trager98} further modify slightly the definitions of
the wavelength bandpasses.  

Since the most recent model grids tabulate ``new'' Lick indices,
we need to consider the effect of
switching between the ``old'' and the ``new'' Lick indices
on determinations of metallicity and age.  Given that GCs
have small internal velocity dispersions ($\sigma_v \le 25$ km/sec)
and that the change in bandpass wavelengths is generally less than
2~\AA,
we expect any resulting differences in measured index strengths to be small.
\cite{puzia02} have calculated
indices on both systems from their spectra of Galactic GCs, and found
maximum differences of 0.2~\AA\ for H$\beta$ and for Mgb, 0.15~\AA\
for Fe5270, and 0.1~\AA\ for Fe5335 and NaD.  
We have done the same for the spectra of 12 Galactic GCs from
\cite{cohen98}, and find similar differences.  There is no
change in any of the indices used here larger than 0.2~\AA,
with the exception of Fe5335 in the most metal rich Galactic GCs.
This is a relatively weak Fe line, in a region where the continuum
is chopped up by many features only slightly weaker (e.g. the Cr~I blend
at 5297~\AA, the blend of Fe~I lines between 5363 and 5371~\AA, etc.)
and hence a small change in the bandpass locations may affect
the value of this
index, and only this index, by more than 0.2~\AA\ out of a total
of about 3~\AA.  

We thus conclude that switching between the various definitions
of the Lick indices does not introduce substantive changes
in the measured indices of a feature for GCs.
For our sample of Galactic GCs, changing from the ``old'' to the
``new'' Lick indices and then deriving their metallicities and ages 
as described in \cite{cohen98} produces a
mean decrease in deduced age of 1 Gyr (the median difference is 0 Gyr).
[Fe/H] is decreased in the mean for such a switch by
0.05 dex; the median change in [Fe/H] is also $-0.05$ dex.

The one published claim that large differences in the measured Lick indices 
are produced
by switching between the ``new'' and ``old'' index bandpass definitions
by \cite{maraston03} (see their Table 1) has been withdrawn;
see the note added in proof in their paper.

\section{Quantitative Interpretation With Models}

A model of the integrated light of a stellar population
can be constructed assuming a star formation history,
a suitable set of stellar evolutionary models and hence isochrones
and a set of integrated spectra as a function of stellar
atmosphere parameters, see e.g. \cite{vazdekis99}.  
These spectra can then be manipulated
to yield parameters that can be compared with observations
of such systems.  For the old stellar systems considered here,
we assume a single burst model with all stars being coeval.

Models for the Lick indices as a function of metallicity
and age such as those of \cite{worthey94b}, used in our
previous work on M87 did not, in fact, integrate
synthetic spectra to determine indices, but rather used fitting functions
based on the Lick group's observations of a grid of calibrating stars which
define the feature strength as a function of effective temperature,
surface gravity and metallicity of a star \citep*[see, e.g.][]{worthey94a}.

These models assumed Solar
abundance ratios, or more precisely, they assumed that the mean
of the calibrating stellar grid observed is equivalent to the
Solar abundance ratios.
High dispersion abundance analyses
of Galactic GCs, however, have revealed that enhancement of the
$\alpha$-elements, specifically Mg and Ca, is almost
universal.  In particular, this is seen in NGC 6528
by \cite{carretta01}, and in NGC 6553 by \cite{cohen99}.
These are two of the most metal rich Galactic GCs known.
High precision analyses of somewhat more
metal poor Galactic GCs also show this.  For example, in M71
\cite{ramirez02} find the mean [Ca/Fe] and the mean [Mg/Fe] 
to be $+0.35\pm0.02$ dex, and in M5 \cite{ramirez03} find
the mean [Mg/Fe] to be $+0.29\pm0.03$ dex while the mean
[Ca/Fe] is $+0.33\pm0.03$ dex.  Similar $\alpha$-enhancements
are found in very metal poor halo field stars, as shown by \cite{cohen02}
and \cite{carretta02}, with references to earlier work by many groups
therein.

Early type galaxies also show enhanced Mg/Fe ratios, with
typical values being about twice Solar \citep{worthey92,
trager00,davies01,proctor02,eisenstein03}.  The origin of
such enhancements in terms of galactic chemical evolution
is well understood; see, for example, the review of
\cite{mcwilliam97}. They are expected to occur
when star formation proceeds rapidly, exhausting the gas prior to the
time when SN Ia begin to spew out Fe and related elements,
whereas the more rapid evolution of massive stars permits 
contributions to the ISM from Type II SN.

\cite{trippico95}  explored the sensitivity
of the Lick indices to changes in the element abundance ratios, while
in the absence of detailed stellar evolutionary grids of models
with non-Solar elemental ratios, \cite{chieffi91} suggested rescaling
of the Solar composition isochrones using the total [Z/H]
to obtain $\alpha$-enhanced isochrones.  Recently stellar evolution
codes and the resulting isochrones  have been computed 
with non-Solar abundance ratios \citep*[see, e.g.][]{vandenberg02}, 
hence calculations which 
predict the Lick indices as a function of Z, $\alpha$/Fe and age 
are becoming available.  We adopt
those of \cite{thomas02}\footnote{\cite{thomas02} 
tabulate ``new'' Lick indices.}.
The metallicity scale of these models,
a subject to which we will return later, is defined in
\cite{maraston03}, where their tests using Galactic GCs are described.
\cite{thomas02} present Lick indices calculated from their models
tabulated at 6 metallicities, [Z/H] = $-2.25$, $-1.35$, $-0.33$,
0.0, +0.35 and +0.67 dex, and at 10 ages ranging from 1 to 15 Gyr.

\begin{figure}
\epsscale{0.9}
\plotone{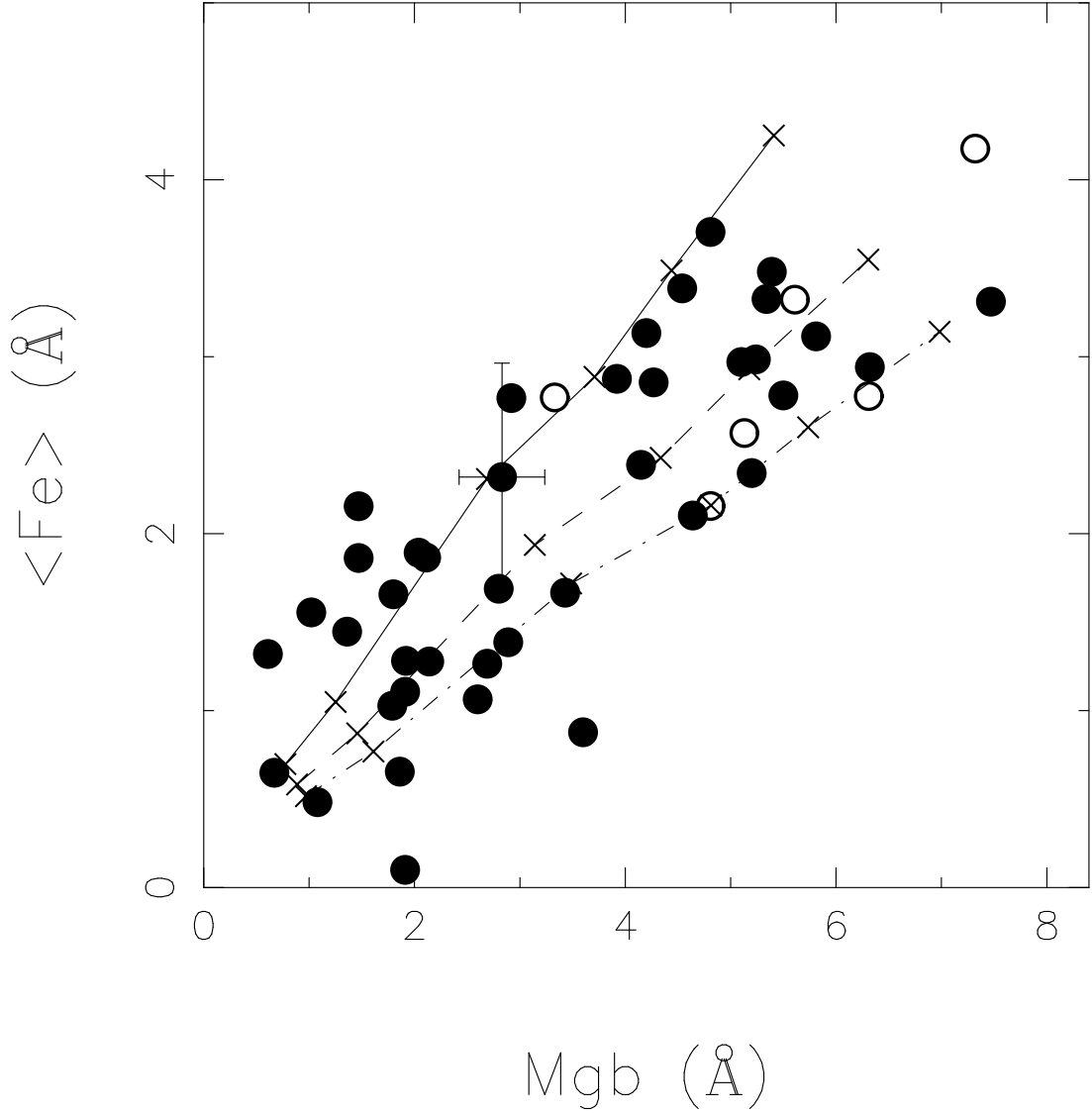}
\caption[]{The average of the Fe5270 and Fe5335 indices
is shown as a function of the Mgb index for the 47
GCs in M49.  Objects with $SNR < 10$ are indicated by open circles.
The 1$\sigma$ uncertainty is shown for Geisler 4168 only;
Figure~\ref{figure_mgfe} shows
the uncertainties for each GC.
The crosses connected by lines
denote the predictions of the grid of models of
\cite{ thomas02} for an age of 12 Gyr at [Z/H] = $-2.25$, $-1.35$, $-0.33$,
0.0, +0.35 and +0.67 dex
for [$\alpha$/Fe] = 0.0 (upper curve, solid line), 
+0.3 (dashed line) and +0.5 dex (lowest curve, dot-dashed line).
\label{figure_mgfemodel}}
\end{figure}

Figure~\ref{figure_mgfemodel} shows the mean of the 
indices covering the strong Fe lines at 5270 and at 
5335~\AA, denoted $<$Fe$>$, as a function of Mgb for the
47 GCs in M49.  The 1$\sigma$ uncertainties for Geisler 4168
only are shown; figure~\ref{figure_mgfe} shows uncertainties
for each of the M49 GCs.
The crosses represent the positions of the models
of \cite{thomas02} with an age of 12 Gyr from  
[Z/H] = $-2.25$ to +0.67 dex.  Line segments connect the 
predictions of the  models at these metallicities
with [$\alpha$/Fe] = 0.0 (Solar), +0.3
and +0.5 dex.  Note that [Z/H] $\approx$ [Fe/H] + 0.94[$\alpha$/Fe]
\citep{thomas02}.  The loci of the
M49 GCs are inconsistent with the Solar $\alpha$/Fe
ratio; a mean enhancement of about a factor of two in $\alpha$/Fe
is required for a reasonable fit to the observations.  This
enhancement, which is
similar to the enhancements found by high dispersion abundance
analyses for the Galactic GCs and to that
required for luminous early type galaxies,
is found over the full range of metallicities within the M49 GC system,
although this is less certain among the most metal poor GCs,
where the separation of indices for the models
between $\alpha$-enhanced and Solar abundance ratios is,
as shown in Figure~\ref{figure_mgfemodel}, smaller.

The range of metallicity of the majority  of the M49 GCs 
as judged from their position on Figure~\ref{figure_mgfemodel}
and the comparison to the \cite{thomas02} models
appears to be  
[Z/H] from $-1.3$ to +0.5 dex, which for $\alpha$/Fe = +0.3 dex,
corresponds to [Fe/H] = $-1.6$ to +0.2 dex.

\begin{figure}
\epsscale{0.9}
\plotone{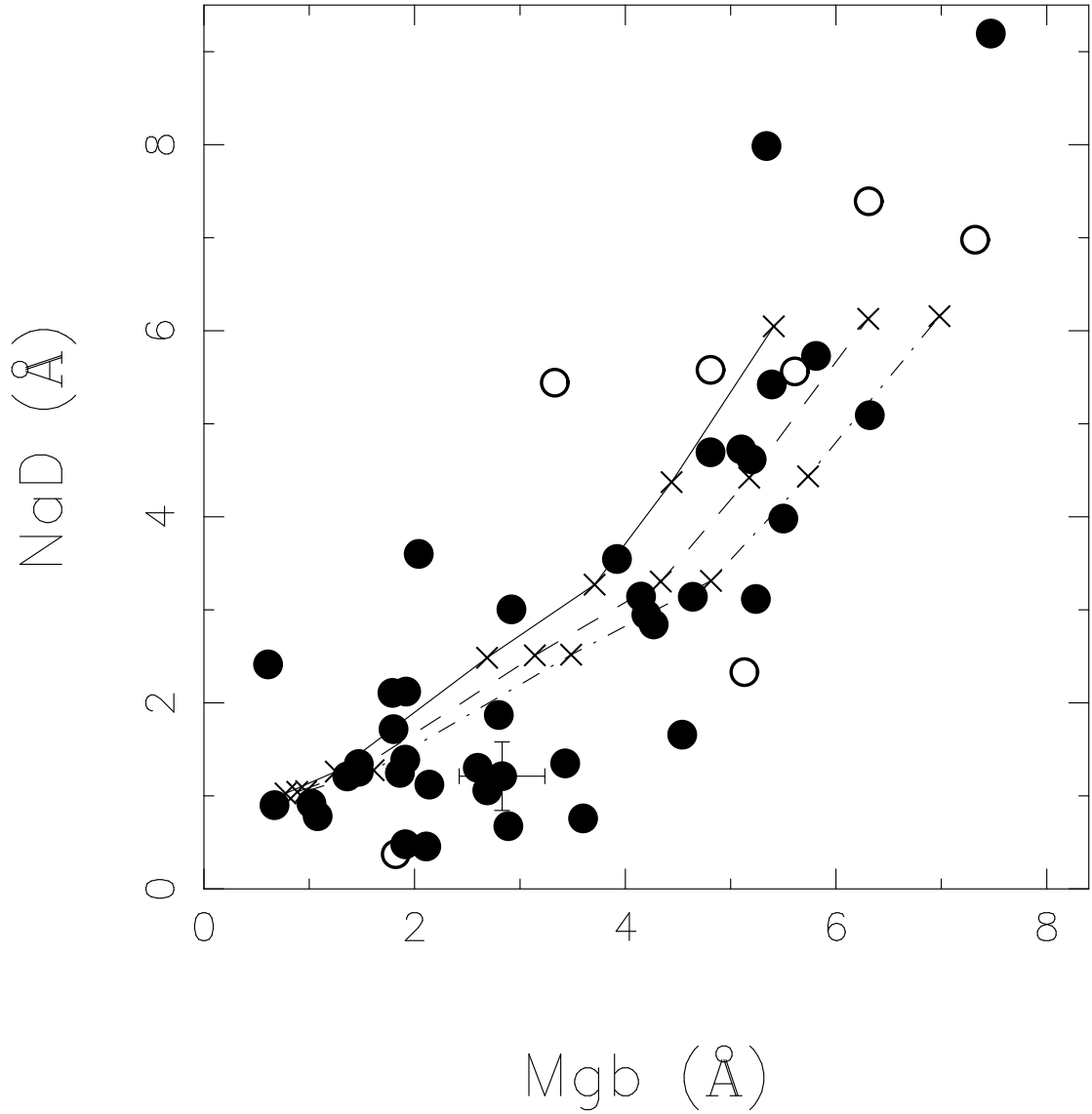}
\caption[]{The NaD index
is shown as a function of the Mgb index for the 47
GCs in M49.  Objects with $SNR < 10$ are indicated by open circles.
The 1$\sigma$ uncertainty is shown for Geisler 4168 only;
Figure~\ref{figure_mgna} shows 
the uncertainties for each GC.
The predictions of  the grid of models of
\cite{ thomas02} for an age of 12 Gyr are shown in a manner
similar to that of Figure~\ref{figure_mgfemodel}.
\label{figure_mgnamodel}}
\end{figure}

Figure~\ref{figure_mgnamodel} shows the NaD index 
as a function of Mgb for the
47 GCs in M49.   Again the 1$\sigma$ errors are shown only for
Geisler 4168; figure~\ref{figure_mgna} shows uncertainties
for each of the M49 GCs.
The crosses represent the positions of the models
of \cite{thomas02} with an age of 12 Gyr with six metallicity values.  
Line segments connect the 
predictions of the models for these metallicities 
with [$\alpha$/Fe] = 0.0 (solar), +0.3 and +0.5 dex.

\begin{figure}
\epsscale{0.9}
\plotone{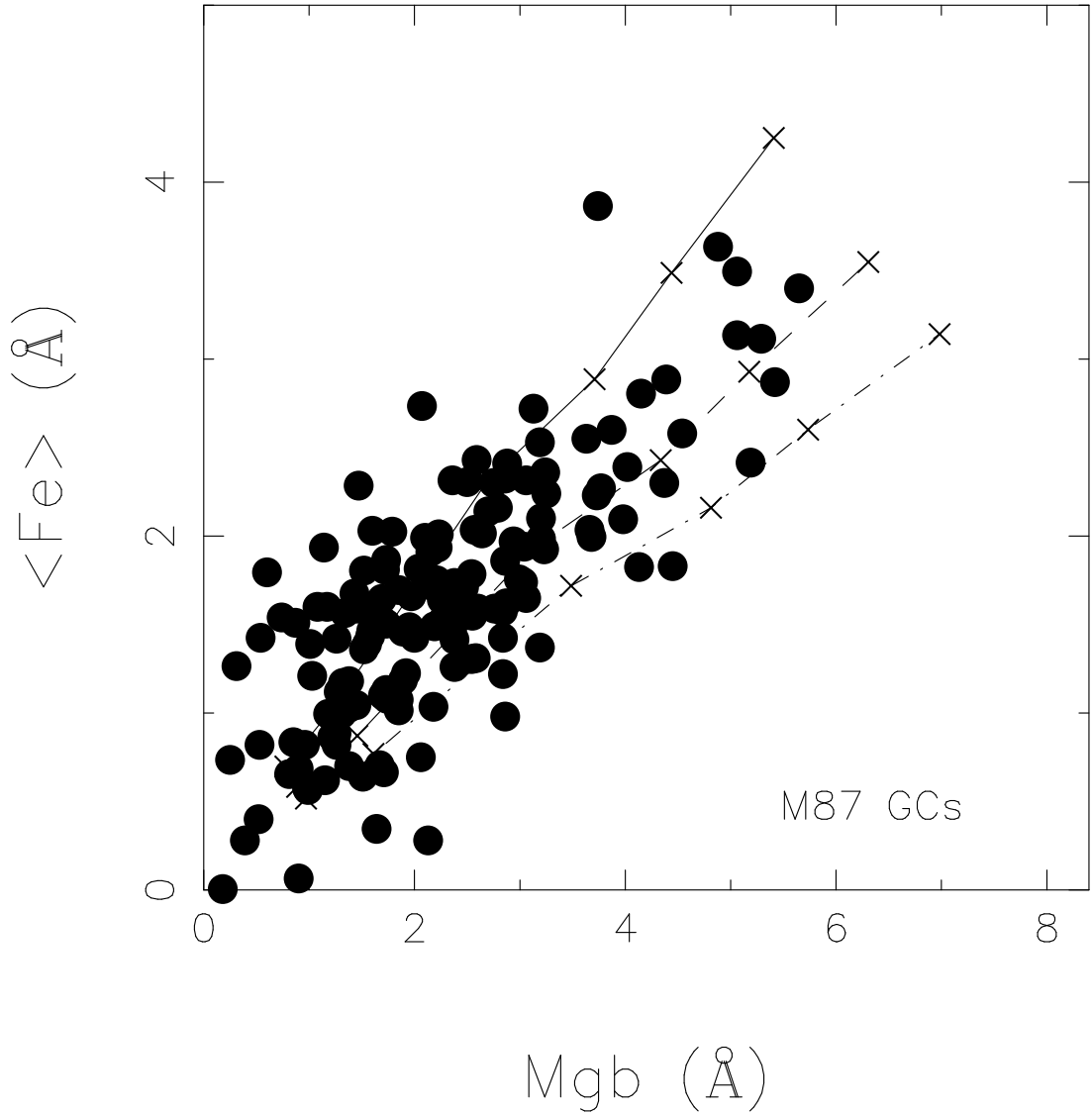}
\caption[]{The average of the Fe5270 and Fe5335 index
is shown as a function of the Mgb index for 150 GCs in M87
from \cite{cohen98}.  
The predictions of  the grid of models of
\cite{ thomas02} for an age of 12 Gyr are shown in a manner
similar to that of Figure~\ref{figure_mgfemodel}.
\label{figure_m87mgfemodel}}
\end{figure}

\begin{figure}
\epsscale{0.9}
\plotone{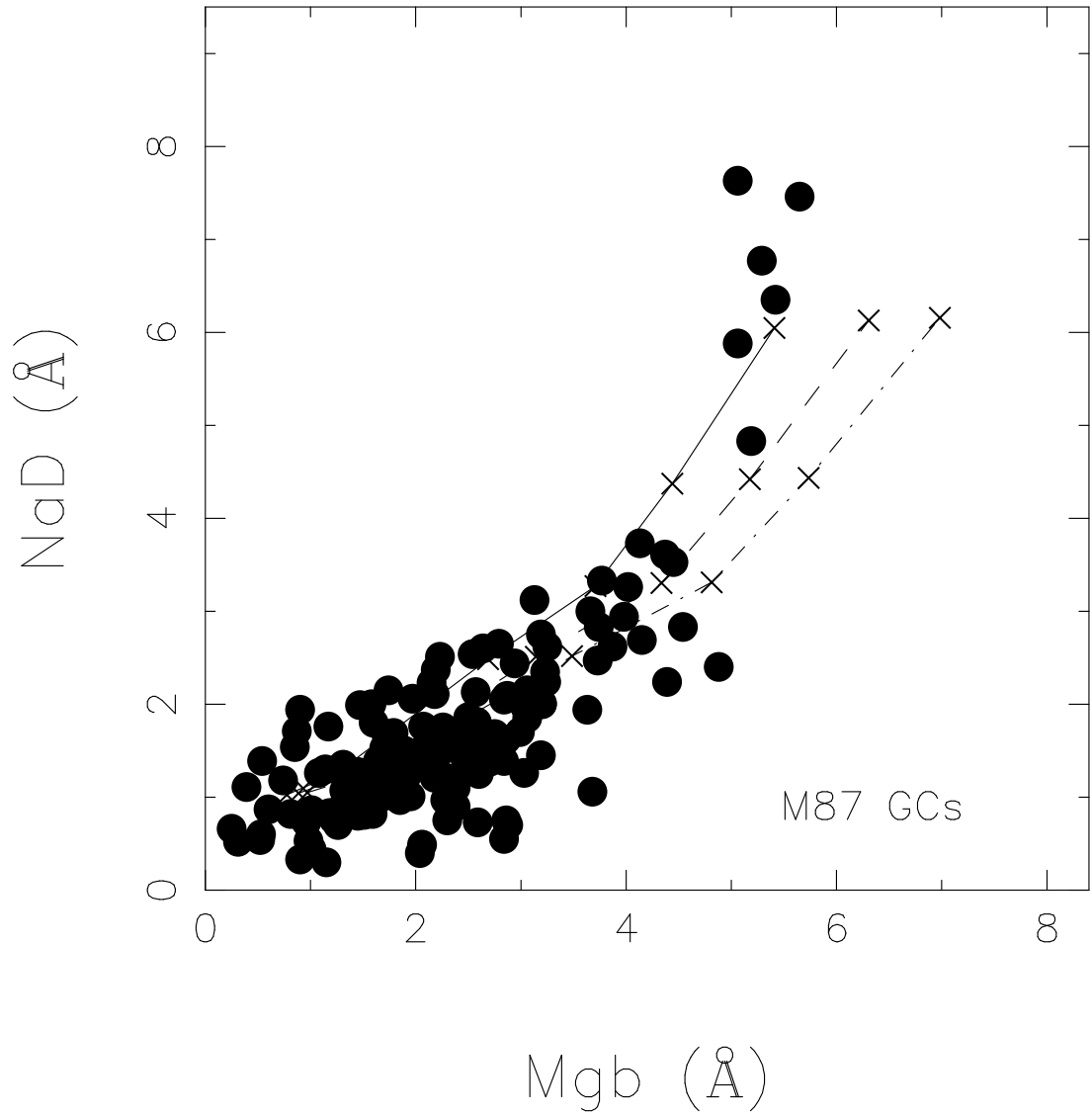}
\caption[]{The NaD index
is shown as a function of the Mgb index for 150 GCs in M87
from \cite{cohen98}.
The predictions of  the grid of models of
\cite{ thomas02} for an age of 12 Gyr are shown in a manner
similar to that of Figure~\ref{figure_mgfemodel}.
\label{figure_m87mgnamodel}}
\end{figure}

In our earlier work on the M87 GC system, the most obvious
shortcoming of
the \cite{worthey94b} models was that, at the metal rich end,
the Fe indices of the M87 GCs
were always smaller for a given Mgb index
than those predicted by the models.  As is shown in 
Figure~\ref{figure_m87mgfemodel},
this problem is now cured by the
use of the new $\alpha$-enhanced models of
\cite{thomas02} for both M87 and for M49.
The major remaining issue, illustrated in 
Figure~\ref{figure_mgnamodel} for M49 and in Figure~\ref{figure_m87mgnamodel}
for M87, is the very strong NaD indices seen in
a few of the most metal rich M49 and M87 GCs,
an issue also noted as problematical in \cite{cohen98}.
We suggest that this is a genuine problem with the models,
which are, as noted above, poorly calibrated for such high
metallicities.

NaD and Mgb are the strongest indices analyzed here.  But the NaD index
is dominated by the
contribution from the sodium doublet, while  the Mgb index 
includes the three strong Mg triplet lines as well as several weaker 
Fe features. Thus the NaD lines should reach the damping part of the
curve of growth at a lower abundance than any of the other features used here.
If the lines are becoming damped for the
majority of the stars contributing to the integrated light,
then line strength  $\propto \sqrt{\rm{X/H}}$, and the line index
will increase rapidly with increasing abundance of the 
appropriate element.  An even more rapid dependence of
index on abundance is expected due to a second effect;
higher abundance clusters have cooler red giant branches, hence
stronger lines. 

The slopes of the predicted indices from 
\cite{thomas02} as a function of [Z/H] suggest that the NaD index 
of the integrated light has in effect reached
the damping part of the curve of growth at [Z/H] near
+0.4 dex, but the Mgb index has not yet done so in their
models even at the highest metallicity included.  We ascribe the
very strong NaD lines seen in the strongest lined
GCs in M49 and in M87 to metallicities near [Mg/H] +0.7 dex
(i.e. [Fe/H] $\sim +0.4$, [Z/H] $\sim +0.7$ dex),  but 
not much larger.

\subsection{Extrapolating the Models \label{section_extrap}}

In the case of both the \cite{worthey94b} and \cite{thomas02} models,
the highest metallicity case tabulated has indices which are weaker
than those of the strongest lined M87 or M49 GCs.  This means
a small extrapolation of the models towards higher metallicity
is required.  For the \cite{worthey94b} models, this was carried
out assuming that all the indices were on the damping part
of the curve of growth.  The derived abundances of 8 M49 GCs were
affected.  In the case of M87, we extrapolated the abundances
of the five GCs for which this is necessary; \cite{cohen98} did not do so.

Only three of the M49 GCs are more metal rich than the most metal
rich $\alpha$-enhanced model tabulated by  \cite{thomas02}.
We extrapolated slightly
beyond the lowest and highest metallicites of the \cite{thomas02} model
grid to $-3.0$ and +1.0 dex, which range includes all the M49 GCs.

\subsection{Calculation of Metallicity for the M49 GCs \label{sec_metal}}

We have established above that the M49 GC indices, including
the key H$\beta$ and H$\alpha$ indices, show similar correlations
to those seen among the Galactic GCs and those of M87 \citep{cohen98}.
We therefore assume an age of 12 Gyr for all of the M49 GCs and
proceed to calculate their metallicities
in two different ways. 
We use the Mgb, NaD and $<$Fe$>$ indices, giving double
weight to the former due to its larger range\footnote{The
NaD index has a similar range, but has problems with
removal of night sky emission.} with comparable accuracy,
with the [$\alpha$/Fe] = +0.3 dex models of \cite{thomas02}.
These produce [Z/H], not [Fe/H].   The resulting [Z/H]
and its 1$\sigma$ error are given  for each M49 GC in the second and third
columns of Table~\ref{table_abund}. 
For the three M49 GCs with the strongest NaD lines, we indicate
in Table~\ref{table_abund} the resulting (lower) abundance
[Z/H] deduced by ignoring the NaD index.  These GCs are all
above the maximum metallicity considered in the \cite{worthey94b} models.

We derive a second metallicity denoted [Fe/H]$_Z$
using the same procedures and codes as in our earlier
work on M87 \citep{cohen98}.  This relies upon the
\cite{worthey94b} models and produces [Fe/H] on a scale
tied to the GC abundances adopted by \cite{zinn84}, denoted
[Fe/H]$_Z$.   These
values are listed in the fourth column of Table~\ref{table_abund}.

\begin{figure}
\epsscale{0.9}
\plotone{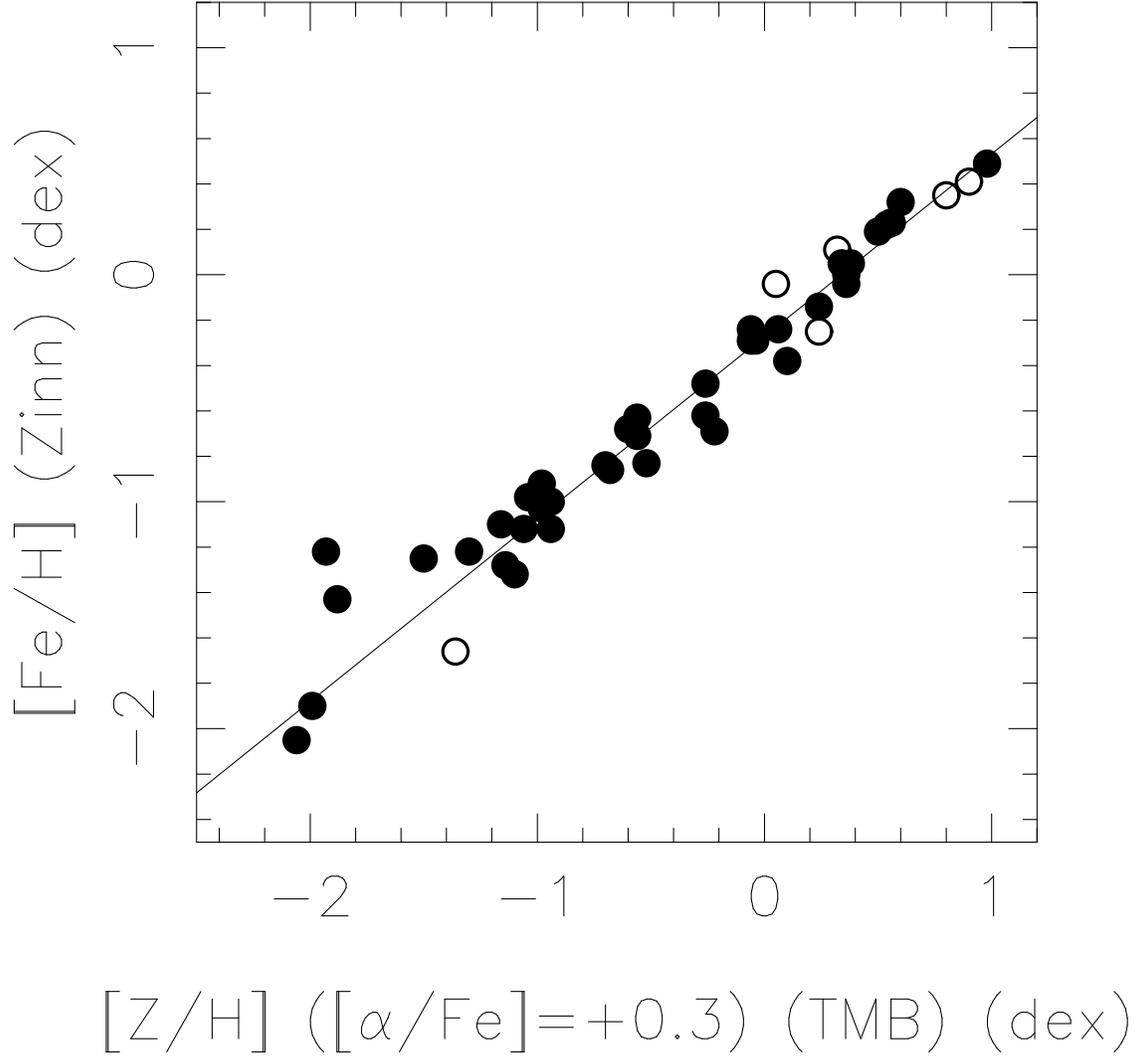}
\caption[]{A comparison of the metallicity of the M49 GC sample
assuming an age of 12 Gyr using two independent sets of models; 
as deduced from the \cite{worthey94b} models set to the
\cite{zinn84} metallicity scale for [Fe/H] versus those derived
from the $\alpha$-enhanced models of \cite{thomas02}.
Objects with $SNR < 10$ are indicated by open circles.
A linear fit is shown.
\label{figure_metalcomp}}
\end{figure}

The two resulting sets of metallicities for the M49 GCs
are shown in Figure~\ref{figure_metalcomp}.  The results
are closely correlated. A linear
fit to the 47 M49 GCs is
\begin{equation} [\rm{Fe/H}]_Z = -0.27\pm0.04  + (0.80\pm0.07)[\rm{Z/H}](TMB),\end{equation} 
with a 1$\sigma$ rms variance about this line of only 0.15 dex.
The origin of this relationship will be discussed in \ref{supersolar}.

\subsection{Comparison of Metallicity Scales \label{supersolar} }

In this paper, we utilize 
three metallicity scales, the one used in M87 by \cite{cohen98},
the one adopted by \cite{geisler96} to analyze their photometry, and
that of \cite{thomas02}.  Although
we used the \cite{worthey94b} models in our M87 analysis, we rescaled
our results to the \cite{zinn84} metallicity scale through calibrating
with Galactic GCs.  The scaling relation derived by \cite{cohen98} was
\begin{equation} [\rm{Fe/H}]_Z = -0.265 + 0.76[\rm{Fe/H}]_W.\end{equation}  
We continue to use this transformation for the M49 GCs, except those
beyond the upper limit of the \cite{worthey94b} model grid, where
it is assumed that [Fe/H]$_Z$ = [Fe/H]$_Z$(max of grid) + $\Delta$,
where $\Delta$ is computed as described in \S\ref{section_extrap}.

\begin{figure}
\epsscale{0.9}
\plotone{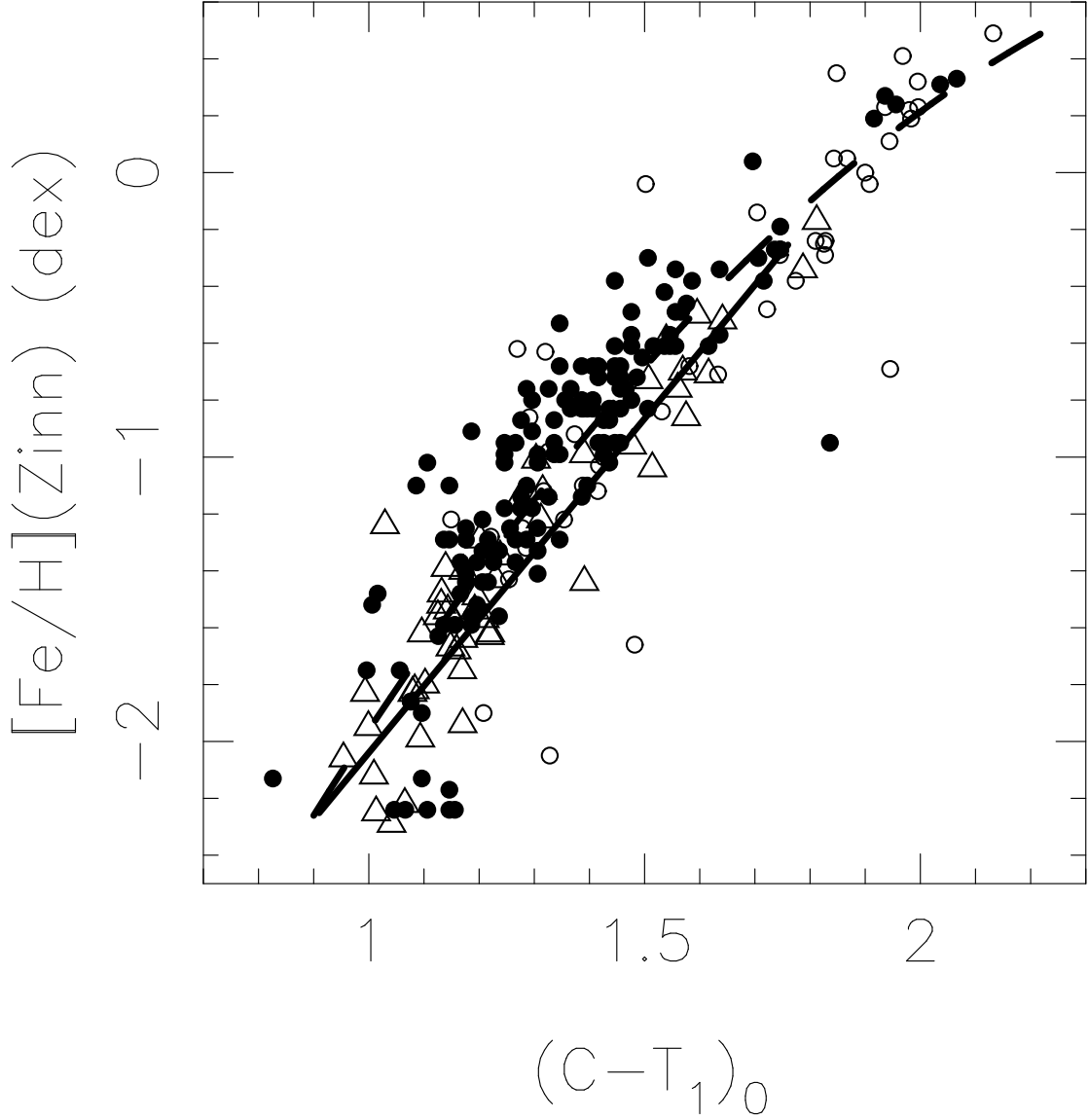}
\caption[]{The dereddened $(C-T_1)$ index of the Washington photometry
is shown as a function of [Fe/H]$_Z$ from \cite{zinn84}
for the Galactic GCs (triangles), and of spectroscopic metallicity for
the M87 (filled circles) and M49 GCs (open circles), 
with sources of data and reddenings given in the text.  The solid line
denotes the relation adopted by \cite{geisler90}.  The dashed line is the
second order fit derived here.
\label{figure_washington_calib}}
\end{figure}

\cite{geisler96} also used the \cite{zinn84}
[Fe/H] scale with a calibration through Washington photometry
of Galactic GCs.  \cite{geisler90} found [Fe/H] = 2.35($C-T_1) -4.39$,
calibrated over the range $-2.25$ to $-0.25$ dex,
which transformation is adopted by \cite{geisler96} throughout
their analyses, even though the colors
of the reddest GCs in M49 exceed the upper limit of the metallicity
calibration.   We are able to present a new calibration of 
$(C-T_1)$ with metallicity, incorporating photometric data for the Galactic
GCs from \cite{harris77} with reddenings from \cite{harris96}, the GCs of M87 
\citep*[data from][]{hanes01} and those of M49 \citep*[data from][]{geisler96},
adopting the reddenings for M87 and for M49 from \cite{schlegel98}.
The metallicities are taken from \cite{zinn84} for the Galactic GCs,
the spectroscopic analyses of
\cite{cohen98} for the M87 GCs, and of the present paper for the M49 GCs.
Figure~\ref{figure_washington_calib} shows the result, i.e. 
[Fe/H]$_Z$ as a function of the dereddened $C-T_1$ color
for this database of 241 GCs.
We find that the linear relationship of \cite{geisler90} is overall
a good fit to the data, except
among the most metal poor GCs, where the actual relationship is steeper
than their adopted fit, as might be
expected since any photometric system will lose sensitivity among the
most metal poor systems.  The second order fit we obtain
after rejecting the 12 most discrepant GCs is
Fe/H]$_Z$ = $-5.64 + 4.438(C-T_1) -0.75(C-T_1)^2$.  The dispersion
of the 228 GCs about this fit is only 0.20 dex.  This second
order fit is only slightly better than the original linear fit
of \cite{geisler90}; both fail to reproduce well the very steep
drop in [Fe/H] at the lowest values of $(C-T_1)$ below 1.1 mag.
Adopting [Fe/H]$_Z$ = 2.27$(C-T_1) -4.21$ for $(C-T_1) > 1.16$ mag and
[Fe/H]$_Z$ = 6.21$(C-T_1) -8.81$ for $(C-T_1) < 1.16$ mag yields
a slightly better fit, particularly at the lowest metallicities.

The galactic GC  metallicity scale of \cite{zinn84} is in turn tied to the
early high and moderate resolution spectroscopic
work on individual stars in Galactic GCs 
of \cite{cohen83} and references therein.  The
high metallicity end of the Galactic GCs, where accurate
calibration data were very limited until quite recently,
remains an area of concern.
It has taken almost 20 years from JGC's initial efforts
to properly calibrate the high metallicity end of the Galactic GCs,
but recently this was carried out through
high dispersion analyses of stars in the very
metal rich galactic GCs NGC N553 and NGC 6528 by
\cite{cohen99} and by \cite{carretta01}.

\begin{figure}
\epsscale{0.9}
\plotone{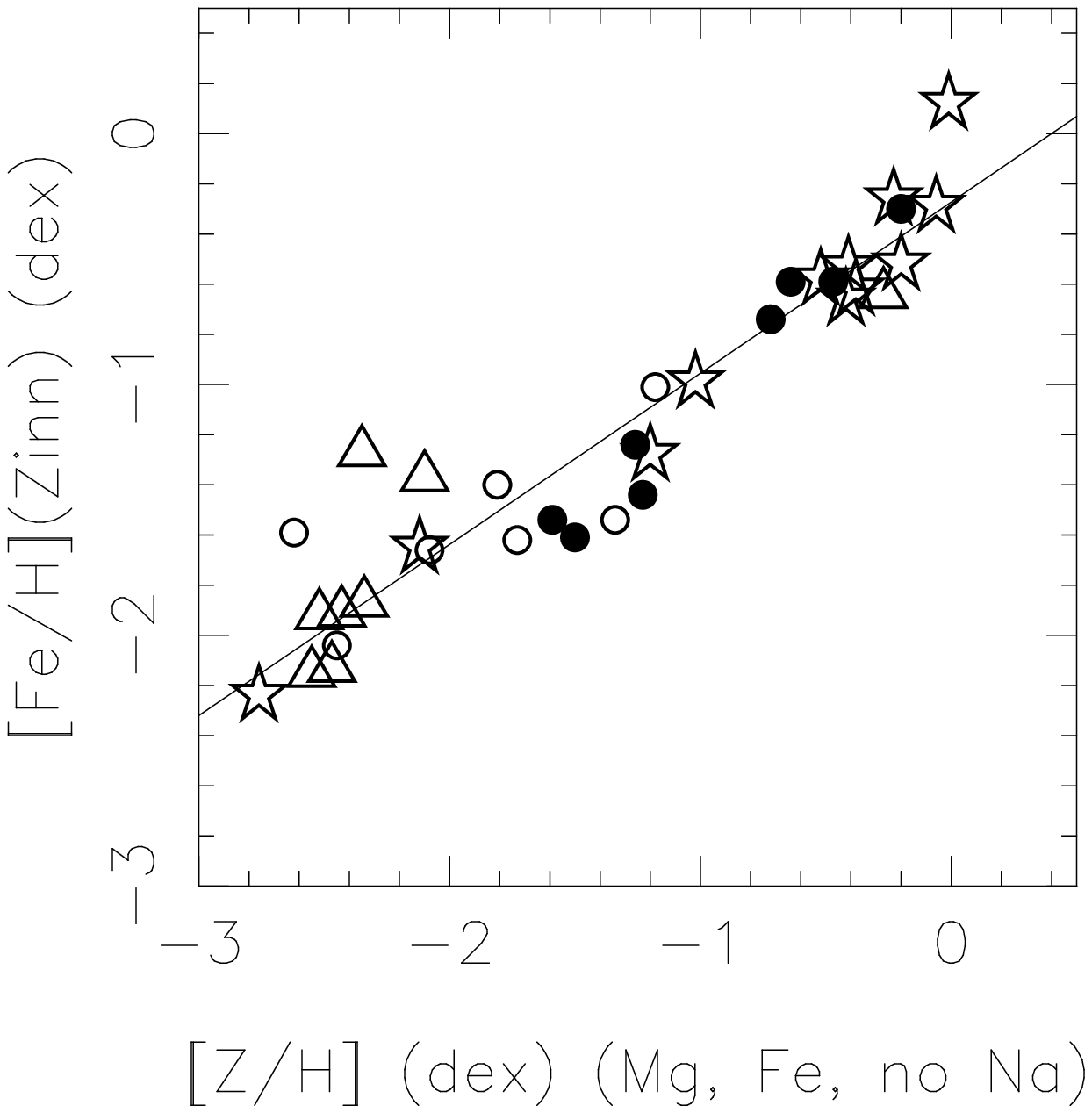}
\caption[]{[Fe/H] from \cite{zinn84} for the 35 
calibrating Galactic globular clusters
are compared to those assigned using the $\alpha$-enhanced models of
\cite{thomas02}.  GCs with data from
\cite{cohen98} are denoted by stars,
those from \cite{puzia02} by filled circles, 
those from \cite{covino95} by triangles and those from \cite{burstein84}
by open circles.  Only the
Mgb and $<$Fe$>$ indices are used for GCs with substantial reddening.
The linear fit is [Fe/H](Zinn) = $-0.27 + 0.68$[Z/H].
\label{figure_galglobmetal}}
\end{figure}

We consider the relationship between the \cite{zinn84}
[Fe/H] determinations for the Galactic GCs and those found
from their Lick indices with the models of \cite{thomas02}.
The NaD lines are ignored when the cluster reddening
is high due to possible contributions from the interstellar medium.
This is shown for the 35 calibrating Galactic GCs in
Figure~\ref{figure_galglobmetal}.  The resulting best linear
fit is 
\begin{equation} [\rm{Fe/H}]_Z = -0.27\pm0.05 + 
(0.68\pm0.03)[\rm{Z/H](TMB)}. \end{equation} 
The 1$\sigma$ rms about this
fit is 0.20 dex.  The major  outliers are NGC 6528, which is 0.25 dex
above the mean line, NGC 7006 and NGC 362.  The first may
actually represent the probable presence of curvature (i.e.
a second order term),
while for the last two, we suggest that the observed indices may be inaccurate.
The constant offset arises from the
$\alpha$-enhancement included in [Z/H].  One would expect the
slope in equation 3 to be unity, but it is not.  In fact, it is very close
to the slopes of equations 1 and 2.
 
\begin{figure}
\epsscale{0.9}
\plotone{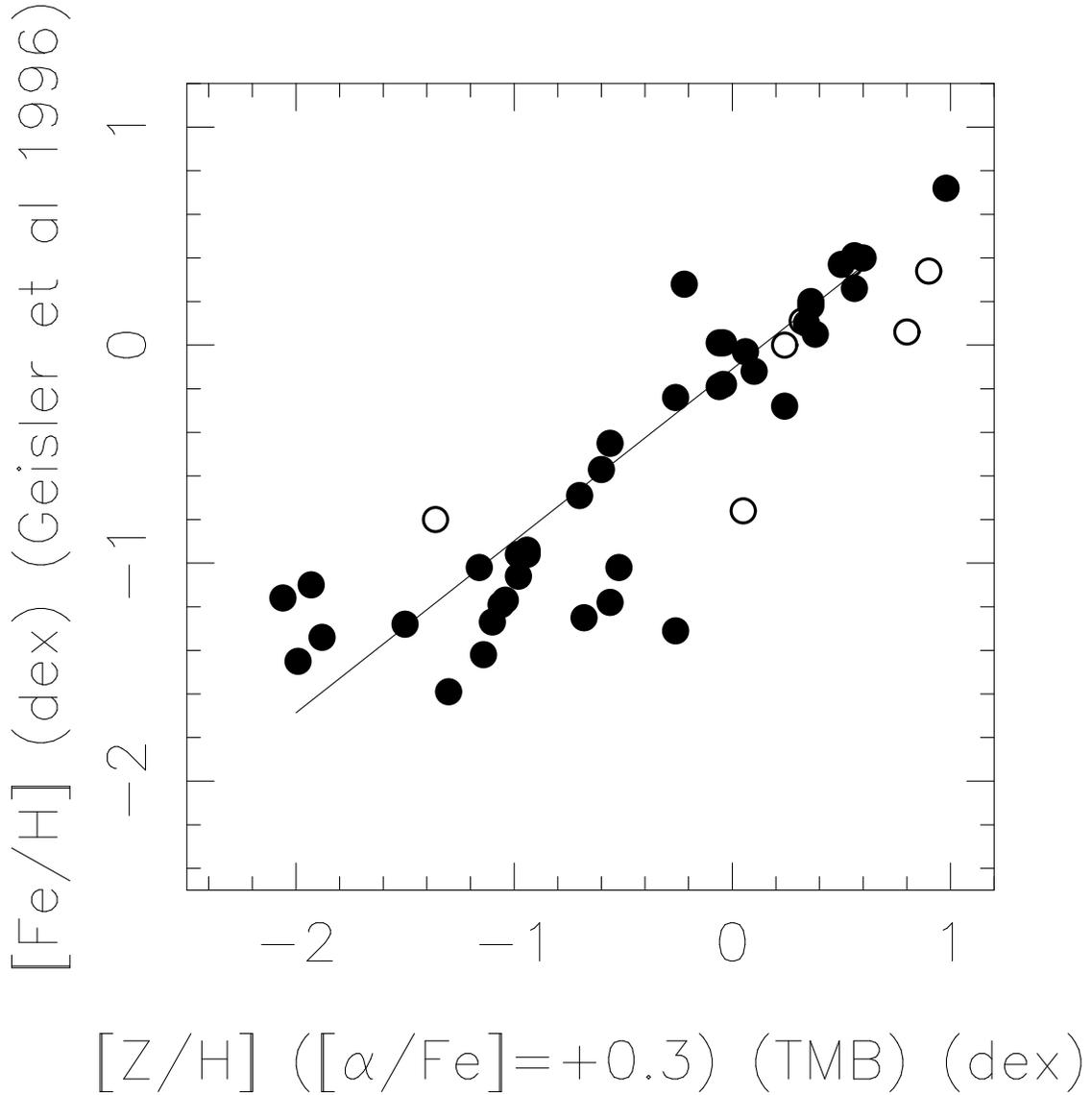}
\caption[]{The metallicities assigned by \cite{geisler96} to the 
M49 GCs from their Washington photometry are compared with
those assigned from LRIS spectroscopy using the
\cite{thomas02} $\alpha$-enhanced model grid assuming a uniform age
of 12 Gyr.
Objects with $SNR < 10$ in their spectra are indicated by open circles.
\label{figure_zphot}}
\end{figure}

We next consider the relationship, shown in
Figure~\ref{figure_zphot}, between the photometric metallicities
for the M49 GCs
derived by \cite{geisler96} and the spectroscopic metallicities
we obtain using the
$\alpha$-enhanced \cite{thomas02} models.  It looks fairly
good except at the low metallicity end.  There any photometric
metallicity indicator is rapidly losing sensitivity, and even
precise photometry cannot easily produce accurate GC metallicities.  
The linear fit shown in the figure is 
\begin{equation}[\rm{Fe/H}](Geisler) = -0.11\pm0.04 + (0.79\pm0.07)[\rm{Z/H}](TMB). \end{equation}  
The dispersion around this fit is 0.31 dex.
Again we see a constant offset due to the use of $\alpha$-enhanced models
by \cite{thomas02} with a slope comparable to those
of equations 1, 2 and 3.

In order to understand the origin of the non-unity slope in equations
1 to 4, we must look carefully at the
three metallicity scales involved.  The abundances of the
most metal rich Galactic GCs are still controversial.
\cite{cohen99} and \cite{carretta01} analyze the spectra of RHB stars 
to derive [Fe/H] = $-$0.06 dex 
and [Fe/H] = +0.07 dex (with an error of $\pm0.15$ dex)
for NGC 6553 and for NGC 6528 respectively.
However, other recent spectroscopic analyses still
obtain much lower values.
We do not consider the work of \cite{coelho01} due to the low dispersion
of their spectra.  The work of \cite{barbuy99}, who found
[Fe/H] = $-0.55\pm0.2$ dex for NGC 6553, is suspect due to the
complexity of the
spectra of the very cool giants near the tip of the RGB.  
\cite{origlia02} obtained, from spectra in the near-IR, [Fe/H] = $-0.3\pm0.2$
dex, which agrees to within 1$\sigma$ with our adopted result.

All of the high dispersion studies mentioned above find that 
these very metal rich
galactic GCs show $\alpha$-enhancements of about a factor of 2. 
It is, however,  much harder to get the absolute abundances, i.e. [Fe/H],
correct than to obtain reasonably accurate abundance ratios.
We adopt here the [Fe/H] values from 
\cite{cohen99} and \cite{carretta01} to define the metal rich end
of the galactic GC calibration.

The values adopted by \cite{zinn84} for NGC 6553 and NGC 6528 are
   [Fe/H]$_Z$ = $-0.29\pm0.11$ and +0.12$\pm0.21$ dex respectively; for
   NGC 6528, this is close to our adopted value, but their abundance
   for NGC 6553 is 0.23 dex lower than ours.  The values for these
two GCs from the \cite{thomas02} models are (ignoring the NaD index)
[Z/H] $= -$0.01 and $-0.06$ dex respectively.  These
values are significantly lower than those we adopt when
the $\alpha$-enhancement is factored in.
The spectroscopic evidence cited above suggests that the
\cite{zinn84} scale is closer to the best data available today than is
the metallicity scale of \cite{thomas02}.
\cite{carretta01}
discuss in detail the calibration of the \cite{zinn84} 
metallicity scale
using the set of the best high dispersion spectroscopic 
analyses of individual stars in Galactic GCs available to date.

Another important part of this problem of the origin of non-unity slopes 
found in equations 1-4 lies at very low metallicities.  
On the basis of the Mgb indices alone, the eight most metal poor
calibrating Galactic GCs, with $-2.24 < $[Fe/H]$_Z < -1.7$ dex,
all fall somewhat below the lower end of the metallicity of 
the \cite{worthey94b} model grid at [Fe/H]$_W = -2.0$ dex
and well below the end of that of
\cite{thomas02} metallicity models at a nominal [Z/H] of $-2.25$ dex.
(In their metallicity calibration, they
only consider Galactic GCs with [Fe/H]$_Z \ge -1.5$.)

Together, these differences in scale 
at the metal-poor and the metal-rich end of the
Galactic GC distribution give rise to the slope of the 
linear relationship between the
Zinn and \cite{thomas02} or \cite{worthey94b} 
metallicity scales being $\sim$0.8 in equations 1 to 4.

An additional and very serious concern is that 
Galactic GCs simply do not span the full range
of metallicity achieved by the M49 and M87 GCs, and so the models for
the highest metallicity range of interest for the Virgo GCs are poorly
calibrated and  poorly tested.
Fitting functions for the Lick indices based on observations
of field stars are not well defined at such high metallicities either.
\cite{worthey94b} provides
data to [Fe/H]$_W$ = +0.5 dex, while \cite{thomas02} calculate indices
to [Z/H] = +0.67 dex.

\section{Metallicity Distribution of the M49 GCs}

\begin{figure}
\epsscale{0.9}
\plotone{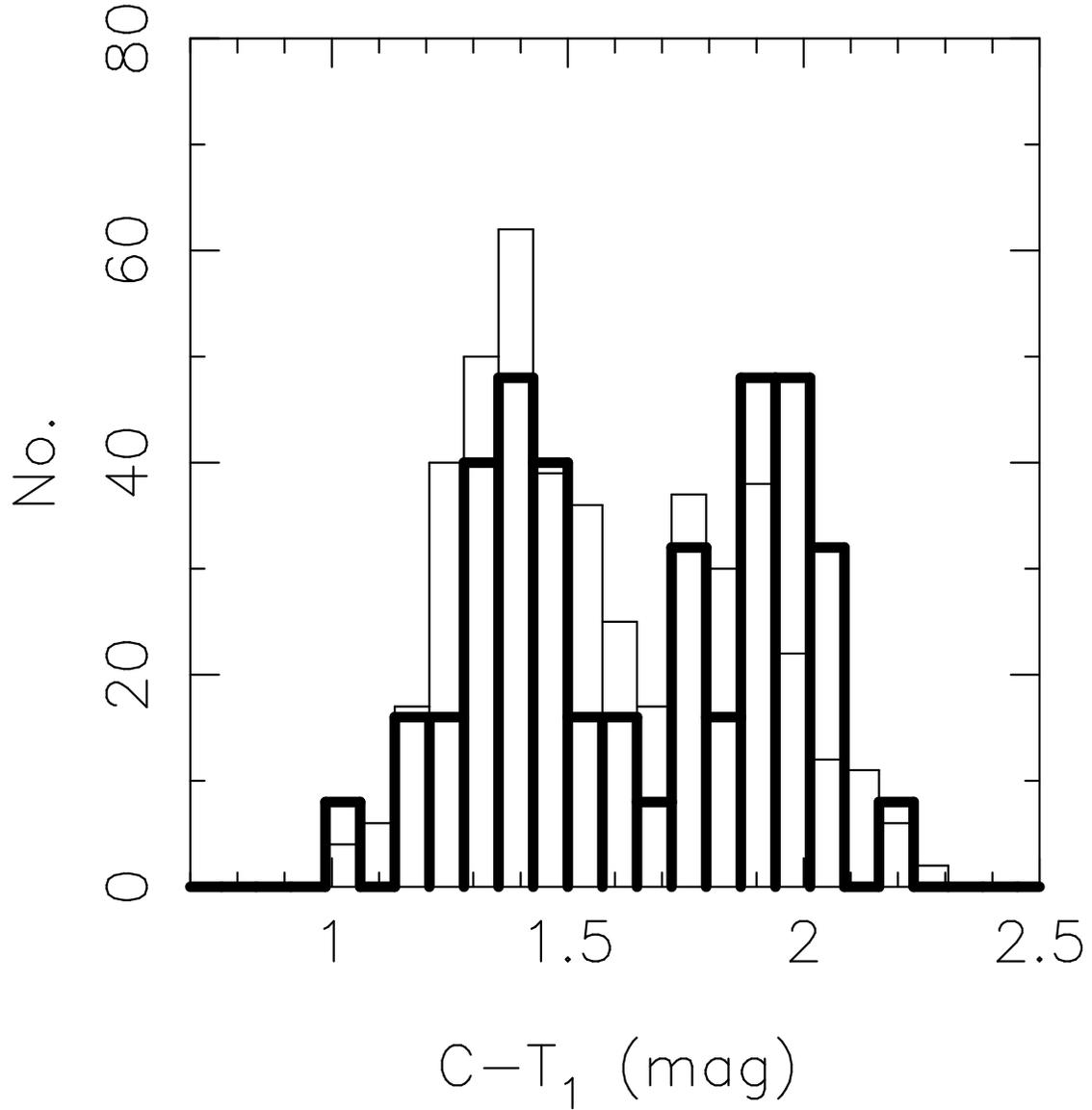}
\caption[]{A histogram of the C-T1 color of the
full sample of candidate GCs in M49
from \cite{geisler96} is shown as the thin line.  A histogram
of this color for the spectroscopically observed sample of GCs is
shown as the thick line.  It is plotted on a vertical scale
1/15 as high as that of the full sample.
\label{figure_hist}}
\end{figure}

We first need to establish whether the spectroscopic sample is
representative of the  M49 GC system.  We assume that 
the complete list of GC candidates of \cite{geisler96}, within
the broad color cut we adopt, is
free of selection effects that are color dependent.
This may not be true, as foreground stars will mostly
be M dwarfs at the red end of the color distribution of the GC
candidates.  Background galaxies
and M dwarfs, however,
appear to be rare interlopers within the central region  of the 
\cite{geisler96} sample.  Figure~\ref{figure_hist} displays
a histogram of the GC candidates extracted from 
photometric database of \cite{geisler96} 
within the radial range of our M49 GC sample and brighter than
our limiting magnitude,
as well as similar histogram for
the sample of M49 GCs observed spectroscopically, including those candidates
which turned out not to be M49 GCs.
(Note that the vertical scale is different for the two samples; that of
the former is eight times larger.)
The spectroscopic sample appears to be a reasonably fair representation
of the M49 GC system. \footnote{A KS test shows that the probability
that the two distributions are identical is low (19\%), but
our goal is a rough estimate of a small number 
(a maximum of two) of parameters
describing the  metallicity distribution
of the M49 GCs.}
Hence the correlations of the indices among the spectroscopic
sample of M49 GCs can be used to delineate the trends
of metallicity and age and related statistical properties within the M49 GC system.
Note in particular that we have sampled close to the full range in 
$C-T_1$ color of the M49 GCs candidates.

The 95\% low metallicity  level given in Table~\ref{table_range} for
the M49 GCs is obtained by sorting the abundances from
Table~\ref{table_abund} \citep*[Table 1 of][for M87]{cohen98},
and is $\sim-2$ dex; the upper end of the metallicity range is
obtained in a similar manner.
The maximum metallicity achieved by the M49 GC system is
[Fe/H]$_Z$ = +0.4 dex, with  $\alpha$/Fe $\sim$ +0.3 dex.

The mean metallicity of the GC system of M49 is substantially
below that of the galaxy itself, for which several 
Lick indices have been measured by \cite{davies93} and by \cite{fisher95}
from long slit spectra.  The highest metallicity reached by
the M49 GCs is, however, comparable to that of the central
region of the galaxy M49 itself,
given as [Fe/H] = +0.24 dex, with [Mg/Fe] = +0.25 dex, by
\cite{terlevich02}.
This is also true for the Galactic GC system.  
M49 is a very luminous galaxy, and a metallicity-luminosity
relation exists among early type galaxies \citep[e.g.][]{kuntschner02},
so it is not surprising to find that
the mean metallicity
of M49 and  of the most metal rich of its GCs to be 
two to three times Solar in [Fe/H], but with a significant
$\alpha$-enhancement.

Table~\ref{table_peaks} lists the total luminosity of the galaxy and
the mean metallicity and the 
approximate location of the peaks in the metallicity distribution
for the five best studied GC systems, M49, M87, NGC~1399
(the central galaxy in the Fornax cluster), M31 and the Galaxy
as determined from ground based and HST photometry and from our
spectroscopy.  These values are all given on the
\cite{zinn84} scale. 
The locations of the two peaks determined from our spectroscopic sample
in M49 is quite uncertain ($\pm0.3$ dex) due to the relatively
small size of the sample. Yet another metallicity calibration is
required here to convert the $V-I$ photometry from HST to
[Fe/H]; we adopt the relationship given by \cite{barmby00}.
The agreement between the photometric and
spectroscopic determinations of the peaks in the M87 GC system
is good. \footnote{The $V-I$ conversion of
\cite{kissler98} when applied to the HST photometry
yields significantly worse agreement with the spectroscopic
metallicities.  The 
\cite{kissler98} metallicity scale may not be properly calibrated
at high metallicities; $\Delta$[Fe/H]/$\Delta(V-I)$ appears
to be too small.}
All three measures agree to within the stated uncertainties
for the lower metallicity peak in M49.

\section{M49 GC Ages}

H$\beta$ is the index among those used here with the highest
sensitivity to cluster age.  H$\beta$ versus strong metal line
equivalent width relations
have been used to attempt to determine ages from 
integrated light spectra for many years 
\citep[see][for an early example]{rabin82}.
However, the separation with age of the metallicity-H$\beta$ contours
of the models for ages
exceeding 5 Gyr is small, and decreases as the metallicity increases.
The observational errors in the H$\beta$ index are fairly large, certainly
larger than the separation of the contours between 8 and 15 Gyr,
and are to first order independent of metallicity.  This
makes the determination of ages quite uncertain.  It is impossible
to determine ages for individual GCs in M49, and we must resort to
median lines, polynomial fits to the index-index relations of 
the GCs as a whole, or summing of the spectra of individual GCs.

In addition, the treatment in the models
of the horizontal branch, whose hotter stars
can contribute strongly to H$\beta$ \citep{lee00},
has improved with time.  Although \cite{worthey94b} included
the HB as a red clump for all clusters, \cite{thomas02} add a HB whose
characteristics depend on the metallicity of the GC.
This change produces a non-negligable change in the predicted
H$\beta$ index in the integrated light of an old very metal poor cluster.
It should also be noted that the \cite{thomas02} models without
$\alpha$-enhancement predict lower Mgb indices than the
\cite{worthey94b} models of the same metallicity, although
they agree in their predictions for the $<$Fe$>$ and NaD indices.
This may be due to the contribution
of these hotter stars at the bluest wavelengths relevant here.

We determined the ages of the M49 GCs using each of the two model grids.  
First, we used
the H$\beta$ observations of the GCs to define a median line in
H$\beta$ versus Mgb.  This was
combined with the \cite{worthey94b} models, just as was done for the M87
GCs in \cite{cohen98}.  The GCs were sorted by metallicity
and binned into groups of 10.  The median age within each group is given
in Table~\ref{table_age}.  The deduced ages are 10 Gyr or older for
all clusters with [Fe/H]$_Z \le -0.8$ dex. 
Since \cite{beasley00} also used the \cite{worthey94b} models, they too
deduced that all the M49 GCs are older than 14$\pm5$ Gyr. 

Geisler
6051 has the strongest observed H$\beta$ index of all the M49 GCs;
it is 2$\sigma$ stronger than expected for an age of 8 Gyr.
Since for a Gaussian distribution we would expect 5\% of the 
sample to lie off the mean line by
2$\sigma$ or more, it is unlikely that this single discrepant
cluster is young.

Splitting the M49 GCs
into metal rich and metal poor groups
at [Fe/H]$_Z$ = $-0.6$ dex, we find (table~\ref{table_age})
mean ages of 13.9$\pm0.8$ 
and 12.8$\pm0.5$ Gyr for each of these respectively using
the models of \cite{worthey94b}. This
is a very small age difference.

\begin{figure}
\epsscale{0.9}
\plotone{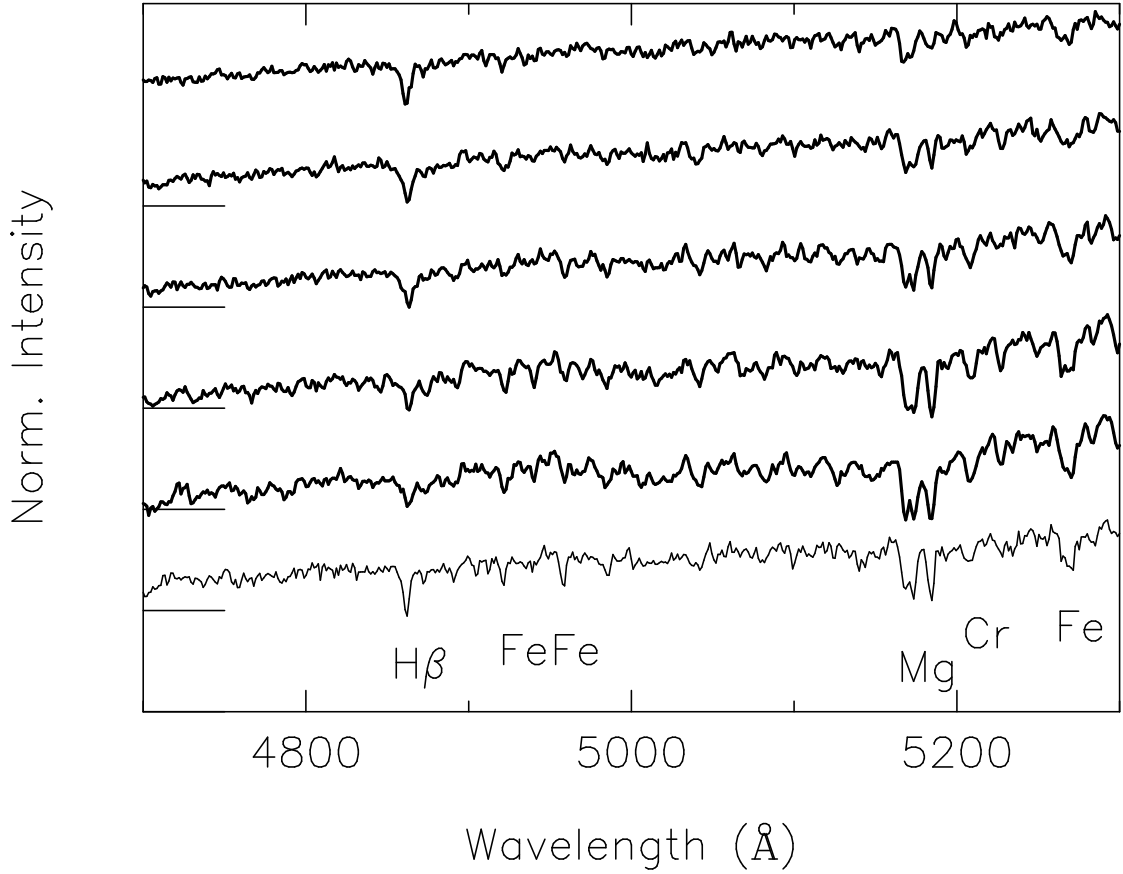}
\caption[]{The sums of 8 spectra per bin 
for five bins of M49 GCs are plotted with metallicity increasing from top to
bottom; the metallicity range of each bin is given in 
Table~\ref{table_tmbage}.
The lowest spectrum displayed (thin line) is that 
of the Galactic star Geisler 2860 ($SNR=29.9$).  
The short horizontal lines at the left edge
together with the bottom of the plot
denote the location of 0 intensity for each spectrum. 
\label{figure_spectra}}
\end{figure}

\begin{figure}
\epsscale{0.9}
\plotone{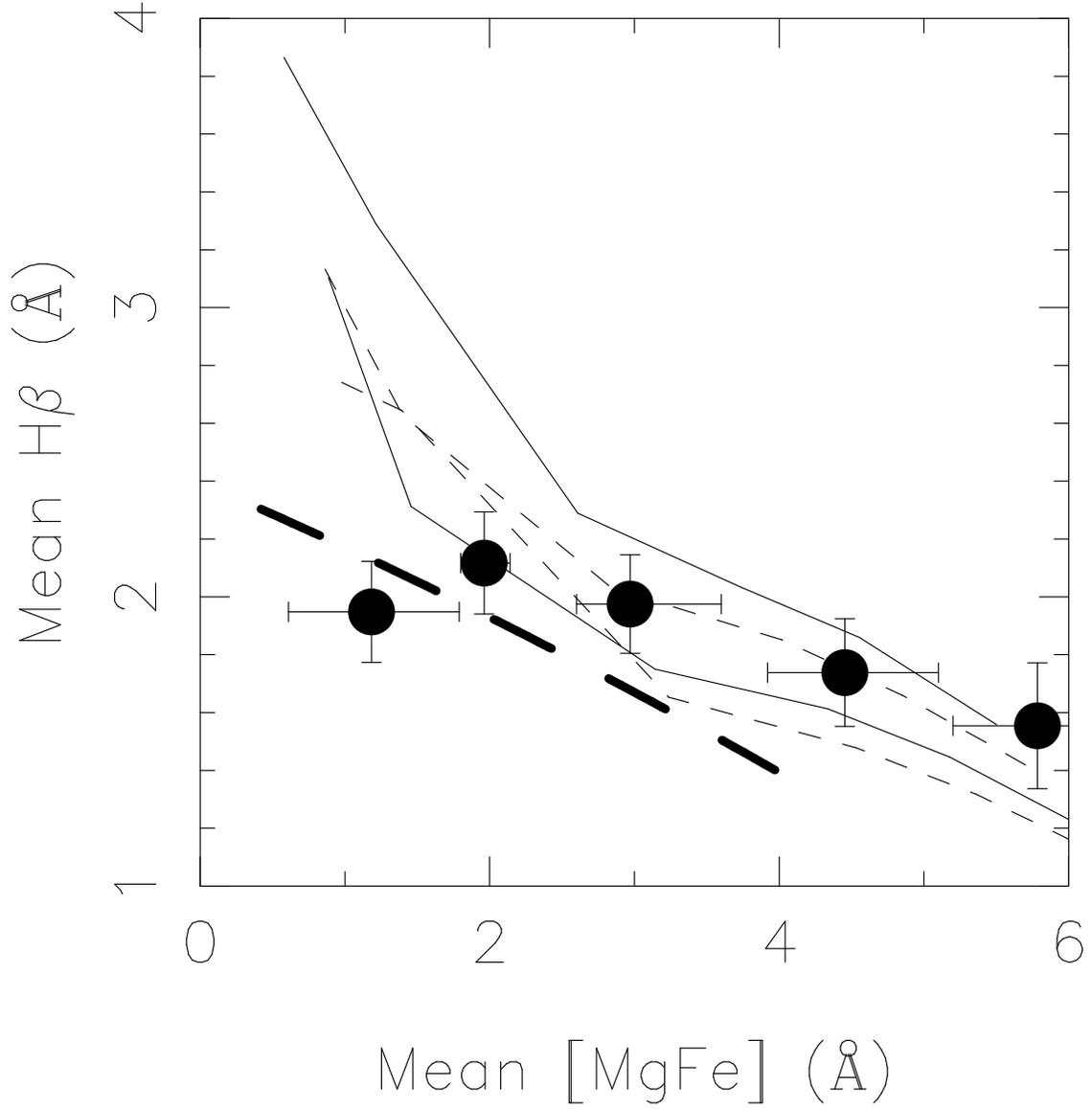}
\caption[]{The H$\beta$ index is shown as a function of the
[MgFe] index for the 40 M49 GCs with $SNR > 10$ sorted in metallicity and
summed into groups of 8.  GCs with low SNR spectra are excluded. 
The X error bar illustrates the range
in [MgFe] for each bin, while the Y 1$\sigma$ error bars are those expected
for the H$\beta$ index of
the summed spectrum of each group.  The thick dashed curve indicates
a second order fit to the sample in M87 of \cite{cohen98}. The models of
\cite{thomas02} are shown for $\alpha$/Fe = +0.3 dex and
for ages of 5 (highest thin curve), 8, 12 and 15 (lowest thin curve) Gyr.
\label{figure_meanage}}
\end{figure}

We also attempted
a similar treatment using the models of \cite{thomas02} of
H$\beta$ as a function of Mgb and as a function of [MgFe]
([MgFe] = $\sqrt{\rm{Mgb<Fe>}}$), which
\cite{puzia03} found to be a metallicity index with minimum 
sensitivity to possible
variations in $\alpha$/Fe.  
We sorted the sample in order of the [Fe/H]$_Z$,
ignoring those with low SNR (SNR $< 10$),
and summed the spectra into 5 bins, each of which contains 8 M49 GCs.
The metallicity range of each bin is given in Table~\ref{table_tmbage}.
The resulting spectra are shown in Figure~\ref{figure_spectra}.
These summed spectra have 8,000 to 10,000 DN/pixel in the continuum
at 5200~\AA\ and each has a minimum SNR of 28.3. 
The figure also shows the spectrum of Geisler 2860,
which is a Galactic star, and whose single spectrum has a SNR of 29.9.
Almost all of the features seen in this figure are real.  Furthermore,
the increase in metallicity from bin to bin and the corresponding
decrease in H$\beta$ line strength are obvious in the summed spectra.
We measured the H$\beta$ index for the summed spectrum of each bin. 
This is shown as a function of [MgFe]\footnote{Here in calculating
[MgFe] we set any negative
Fe5335 or Fe5270 indices to 0.3~\AA.} 
in Figure~\ref{figure_meanage}.  The thin lines
represent the models of \cite{thomas02} for ages of 5, 8, 12, and 15 Gyr.  
The thick dashed line is the second order fit to the M87 GCs.

Table~\ref{table_tmbage} gives the result.  If the models
of \cite{thomas02}
are correct, the ages of the M49 GCs systematically decrease
as the metallicity increases from a mean age of more than 15 Gyr
for the most metal poor GCs to a mean age of 5 Gyr for the most
metal rich, with a mean age for the GC system as a whole of $\sim$10 Gyr.  
Furthermore, the M87 GCs then have ages exceeding
15 Gyr for all metallicities (see Figure~\ref{figure_meanage}).  

We look again at the Galactic globular clusters for clues
as to which set of models for the key H$\beta$ index are better. 
The very strong H$\beta$ indices predicted for the most metal poor
GCs for {\it{any}} age by the \cite{thomas02} models
are not present in the M49 GC sample, where the mean H$\beta$
for the most metal-poor bin is 1.95~\AA\ and the deduced age
exceeds 15 Gyr.  
The \cite{worthey94b} models predict H$\beta$ to be 2.44~\AA\ for very
low metallicity and an age of 12 Gyr, which is a better
fit to our observations in M49 and to those of
\cite{cohen98} in M87, this difference presumably arising
from the different treatment of the HB between the two sets of models.
Evidence from the Galactic globular clusters (see 
Figure~\ref{figure_quadgalglobs}) suggests an intermediate
value for the mean H$\beta$ index among
the most metal poor GCs, but still not as high as the \cite{thomas02}
models suggest.  The scatter of H$\beta$ measured among the
low metallicity Galactic globular clusters, shown in 
Figure~\ref{figure_quadgalglobs}, is large.  If the uncertainties given
by \cite{covino95} are realistic, then this range is too large
to be due to measurement errors, and may represent
variations in the relative contribution to the integrated light
of the HB or variations in the distribution of stars along the HB.

If one wishes to overinterpret
the data and if one believes that the \cite{thomas02} models
better represent reality for the Balmer lines, which they unquestionably do for
the metallic lines due to incorporation of $\alpha$-enhancement, 
one could claim that the most metal poor GCs in M49
are the oldest, and the mean age of the M49 GCs decreases
from 15 Gyr to 5 Gyr as the metallicity increases. For present purposes,
we adopt the age results from the \cite{worthey94b} models.  However, the only
safe statement in our view is that the M49 GCs
are in the mean older than 10 Gyr.
Further speculation about the details of the age distribution of the
M49 GCs requires observational data of
still higher quality as well as an exquisite understanding of
the nature of the HB in extragalactic GC systems, 
neither of which is currently available.

\section{Implications for Globular Cluster Formation}

We have
discovered here that enhancement of the $\alpha$ elements
(Mg, Ca, Na) is common in the GC systems of massive galaxies.  
Following upon the
work of \cite{cohen98}, we have detected this in both M49 and in
M87.  Such an enhancement is a well known feature of Galactic
globular clusters, and those few which are known or suspected
to be young in the Galaxy are also known to have no or smaller
$\alpha$-enhancements \citep{brown97, smith02}. \cite{cohen03}
also has found that a similar $\alpha$-enhancement is 
characteristic of the M31 GCs.

In the Galaxy, where detailed study with exquisite precision
of at least nearby individual stellar populations
is possible, \cite{reddy02} have demonstrated that the stars of the thin
disk in the Solar neighborhood (i.e. with $R_{GC}$ between 7 and 10 kpc)
show very small scatter in element ratios, with element ratios essentially
Solar.  Separating the sample into thick disk stars and thin disk
stars on the basis of their kinematics, they find the former to have in 
the mean higher  enhancements for the $\alpha$-elements Mg, Si, Ca and Ti
than do the thin disk stars.   Furthermore, they find a relatively
clean age-metallicity relationship characterized by a slow
monotonic decrease of metallicity with age, with considerable
scatter about this trend. 

A model for the chemical evolution of a galaxy requires adding
up the contributions of SNII, SNIa and AGB stars. 
If the contribution of type Ia SN, which produce a smaller ratio
of $\alpha$/Fe elements in their ejecta, is delayed compared to
that of SNII, which produce copious amounts of $\alpha$ elements,
but little Fe, $\alpha$-enhancements
can result.  Such circumstances prevailed in the young galaxy, before
SNIa, with their longer required timescale, started exploding, while
SNII come from massive stars with short lifetimes.

The roughly equal $\alpha$-enhancement seen in these four GC systems,
and the fact that this is seen among both metal rich and metal poor
GCs within these systems, provides a crude chronometer and suggests
that the GCs in these systems are mostly old, with ages of
10 Gyr or longer.  Additional recent
spectroscopic studies with much smaller samples
of GCs in quiescent massive bulge dominated galaxies 
including the Sombrero galaxy (M104=NGC 4594) by \cite{larsen02}
and NGC 1399, the central galaxy in the Fornax cluster, 
by \cite{forbes02} find similar old ages.

Relative ages for subsamples within a GC system can be obtained
with a number of assumptions about the GC mass function from
photometric analyses by measuring the
brightness of the turnover of the GC luminosity function
and comparing to predictions from single-burst population models.
This technique has been applied to M49 by \cite{puzia99}, who
found no difference in age larger than 3 Gyr between the metal poor and metal
rich GC populations.   \cite{jordan02} 
have applied such techniques to M87, where again no age difference
of more than 0.5 Gyr
was found  between the metal rich and metal poor populations.
Thus the uniformly old ages found through spectroscopic observations in 
M87 and in M49 when the \cite{worthey94b} models
are used are supported by photometric analyses.
The old ages of the GCs in these galaxies, and the small difference in
their relative ages, are consistent with the predictions of empirical
models for the hierarchical growth that involves no GC formation 
\citep*[e.g.,][]{cote98}.
However, the very steep protogalactic mass
spectra required to explain the detailed shape of the GC metallicity
distributions in the context of this scenario remains problematical

Models for GC formation which involve the formation of GCs in
late epoch
mergers such as those of \cite{ashman92} and \cite{bekki02}
must somehow push
those mergers back earlier in time, earlier than $z\sim2$. 
Evidence from the study of distant galaxies in situ at such redshifts
also suggests that massive galaxies are more or less
completely assembled by $z\sim1$ \citep{cohen02a} and are
at least partially assembled without major ongoing
star formation at $z\sim2$ \citep{labbe03}. 
\cite{beasley02} investigate the formation of GCs within the framework
of the treatment of semi-analytical galaxy formation of \cite{cole00}.
They find age differences of 1--7 Gyr
between the metal poor GCs, which are formed at high redshift in
proto-galactic fragments, and the metal rich GCs, which are formed
during subsequent mergers. The inferred age differences, however,
ultimately depend on their assumption of a truncated formation history
for the metal-poor cluster and a metallicity-dependent cluster
formation efficiency. It appears that fully successful model of GC
formation does not exist at the present time.

\acknowledgements
The entire Keck/HIRES user community owes a huge debt
to Jerry Nelson, Gerry Smith, Steve Vogt, and many other people who have
worked to make the Keck Telescope and HIRES a reality and to
operate and maintain the Keck Observatory.  
We are grateful to the W. M. Keck Foundation for the vision to 
fund the construction of the W. M. Keck Observatory. 
The authors wish to extend special thanks to those of Hawaiian ancestry
on whose sacred mountain we are privileged to be guests. 
Without their generous hospitality, none of the observations presented
herein would have been possible.  We thank the referee for
helpful suggestions.

The extragalactic work of JGC is not supported by any federal agency.
JPB acknowledges support from NASA grant NAG5-7697.
PC acknowledges support provided by the Sherman M. Fairchild Foundation
during the early stages of this work, and additional support provided 
by NASA LTSA grant NAG5-11714.


\begin{deluxetable}{lrrrr}
\tablenum{1}
\tablewidth{0pt}
\small
\tablecaption{Background Galaxies \label{table_gals}}
\tablehead{\colhead{ID\tablenotemark{a}} & 
\colhead{RA} & \colhead{Dec} & \colhead{$R$} & \colhead{$z$} \\
\colhead{} & \colhead{(1950)} & \colhead{(1950)} &
\colhead{(mag)} & \colhead{} }
\startdata 
(1650)  &  12 27 23.68  &  +8 13 42.3 & $>$23.5 & 0.629\tablenotemark{b} \\
1982    &  12 27 09.13   & +8 13 42.1 & 20.6 & 0.086  \\
(1982)  &  12 27 09.97 & +8 13  42.3 & 23.7 & 0.434 \\
(2569)  &  12 27 11.68 & +8 14  31.6 & 23.3 & 0.091 \\
(1798)  &  12 27 12.48 & +8 13  25.0 & 22.1 &  0.314\tablenotemark{c} \\
\enddata
\tablenotetext{a}{IDs in parentheses denote serendipitous galaxies
found in the slitlet intended for the GC candidate listed.  The database
of \cite{geisler96} is used.}
\tablenotetext{b}{This redshift is based on a single emission line assumed
to be that of [OII] at 3727~\AA.}
\tablenotetext{c}{The redshift for this galaxy comes from H+K absorption.
For all the other galaxies in this table, it is derived from emission lines.}
\end{deluxetable}

\begin{deluxetable}{lrrrrrrrrrrr}
\tablenum{2}
\tablewidth{0pt}
\small
\tablecaption{Measured Indices \label{table_indices}}
\tablehead{\colhead{ID\tablenotemark{a}} & \colhead{H$\beta$} &
\colhead{Mg2} & \colhead{Mg1} & \colhead{Mgb} & \colhead{Fe5270} &
\colhead{Fe5335} & \colhead{NaD} & \colhead{Ti01} & \colhead{Ti02}
& \colhead{H$\alpha$} & \colhead{SNR\tablenotemark{b}} \\
\colhead{} & \colhead{(\AA)} & \colhead{(mag)} & \colhead{(mag)} & \colhead{(\AA)} &
\colhead{(\AA)} & \colhead{(\AA)} & \colhead{(\AA)}  
& \colhead{(mag)} & \colhead{(mag)} & \colhead{(\AA)} & \colhead{~~}    }
\startdata 

 1475 &     2.13 &    0.14 &    0.045 &    2.11 &    3.14 &    0.59 &    0.46 &    0.031 &    0.060 &    2.04 & 16.3 \\ 
 1508 &  0.30 &    0.31 &    0.081 &    4.81 &    2.60  &    1.71 &    5.58  &    0.055 &    0.201 &    2.25 & 5.4 \\
 1650 &     1.31 &    0.31 &    0.101 &    5.50 &    2.65 &    2.91 &    3.98 &    0.038 &    0.058 &    0.74 & 15.1 \\ 
 1731 &     1.60 &    0.24 &    0.101 &    4.54 &    2.57 &    4.20 &    1.66 &    0.039 &    0.106 &    1.62 & 11.8 \\ 
 1798 &     1.33 &    0.36 &    0.117 &    5.81 &    2.78 &    3.45 &    5.73 &    0.044 &    0.095 &    0.81 & 14.7 \\ 
 1846 &     0.03 &    0.37 &    0.201 &    5.61 &    3.36 &    3.28 &    5.56 &    0.084 &    0.109 &    1.97 & 8.7 \\ 
 1889 &  2.42 &    0.03 &   $-$0.010 &    0.67 &    1.00 &    0.30 &    0.90 &   $-$0.014 &   $-$0.084 &    2.60 & 14.1 \\
 1892 &     0.89 &    0.07 &    0.018 &    1.82 &   $-$0.03 &   $-$0.55 &    0.37 &    0.052 &    0.021 &    2.48 & 9.4  \\ 
 1905 &  2.85 &    0.08 &    0.015 &    2.04 &    2.58 &    1.20 &    3.60 &   $-$0.007 &   $-$0.004 &    3.25 &  13.4 \\ 
 2013 &     2.47 &    0.08 &    0.044 &    0.61 &    1.50 &    1.14 &    2.41 &    0.030 &    0.043 &    2.35 & 13.2 \\ 
 2031 &     2.87 &    0.10 &    0.027 &    1.80 &    2.07 &    1.25 &    1.72 &   $-$0.012 &    0.010 &   2.19 & 17.2 \\ 
 2045 &     1.58 &    0.19 &    0.105 &    2.92 &    3.16 &    2.38 &    3.00 &    0.034 &    0.063 &    1.95 & 15.2 \\ 
 2060 &     1.74 &    0.08 &    0.017 &    1.02 &    1.33 &    1.78 &    0.92 &    0.006 &   $-$0.159 &    2.63 & 27.6 \\ 
 2178 &     1.59 &    0.04 &    0.010 &    1.86 &   $-$0.04 &    1.35 &    1.25 &   $-$0.013 &    0.024 &    2.68 & 10.8 \\ 
 2188 &     0.90 &    0.13 &    0.027 &    2.60 &    1.75 &    0.37 &    1.30 &    0.025 &    0.040 &    2.03 & 13.1 \\ 
 2306 &     1.86 &    0.18 &    0.087 &    2.80 &    2.74 &    0.64 &    1.87 &    0.034 &    0.048 &    2.24 & 19.9 \\ 
 2406 &     0.66 &    0.35 &    0.157 &    5.39 &    3.96 &    3.01 &    5.42 &    0.052 &    0.092 &   \nodata & 13.6 \\ 
 2421 &     2.33 &    0.13 &    0.011 &    1.47 &    3.55 &    0.76 &    1.26 &    0.018 &   $-$0.027 &    2.17 & 14.3 \\ 
 2502 &  2.41 &    0.23 &    0.098 &    3.92 &    3.50 &    2.25 &    3.54 &    0.035 &    0.048 &   \nodata & 12.7 \\ 
 2528 &     2.33 &    0.09 &    0.038 &    2.14 &    1.57 &    0.98 &    1.12 &    0.012 &    0.039 &   2.00 & 22.0 \\ 
 2543 &     2.13 &    0.08 &    0.020 &    1.47 &    2.21 &    1.51 &    1.34 &   $-$0.010 &    0.029 & \nodata  & 25.6 \\ 
 2569 &     1.75 &    0.29 &    0.131 &    5.10 &    3.30 &    2.64 &    4.72 &    0.039 &    0.063 &   \nodata & 23.3 \\ 
 2813 &     2.16 &    0.30 &    0.168 &    5.20 &    2.57 &    2.12 &    4.62 &    0.016 &    0.078 &  \nodata & 10.4 \\ 
 3150 &     1.78 &    0.26 &    0.112 &    4.15 &    2.74 &    2.04 &    3.14 &    0.020 &    0.119 &    1.49 & 13.0 \\ 
 3603 &     2.14 &    0.26 &    0.098 &    5.24 &    3.27 &    2.70 &    3.12 &    0.037 &    0.083 &    2.05 & 12.8 \\ 
 3788 &     2.61 &    0.18 &    0.086 &    5.13 &    4.64 &    0.50 &    2.33 &    0.018 &    0.069 &    3.18 & 8.5 \\ 
 3900 &     0.64 &    0.43 &    0.177 &    6.31 &    1.87 &    3.69 &    7.39 &    0.018 &    0.083 &    1.24 & 8.5 \\ 
 4017 &     1.37 &    0.08 &    0.037 &    1.92 &    1.00 &    1.56 &    2.12 &   $-$0.002 &    0.011 &  1.20 & 17.1 \\ 
 4062 &     0.14 &    0.37 &    0.106 &    7.32 &    2.98 &    5.38 &    6.98 &   $-$0.018 &  0.080 &    2.08 & 6.9 \\ 
 4144 &     2.23 &    0.04 &    0.009 &    1.91 &    1.44 &    0.78 &    0.48 &    0.011 &    0.015 &    2.31 & 19.7 \\ 
 4168 &     2.14 &    0.17 &    0.067 &    2.83 &    2.66 &    1.98 &    1.21 &    0.022 &    0.060 &    1.49 & 21.6 \\ 
 4217 &     2.13 &    0.26 &    0.062 &    4.20 &    3.96 &    2.31 &    2.95 &    0.034 &    0.037 &    2.85 & 11.1 \\ 
 4296 &     2.14 &    0.05 &    0.012 &    1.91 &    0.86 &   $-$0.66 &    1.39 &   $-$0.018 &    0.004 &    2.47 & 12.7 \\ 
 4351 &  1.54 &    0.05 &   $-$0.021 &    1.08 &    0.97 &   0.00 &    0.78 &    0.031 &   $-$0.016 &  2.20 & 12.5 \\
 4401 &     1.64 &    0.30 &    0.115 &    4.81 &    3.12 &    4.29 &    4.69 &    0.043 &    0.075 &    2.09 & 14.2 \\ 
 4513 &     1.79 &    0.27 &    0.139 &    4.64 &    2.19 &    2.02 &    3.14 &    0.043 &    0.083 &    1.08 & 22.1 \\ 
 4541 &     2.34 &    0.09 &    0.020 &    2.69 &    1.48 &    1.05 &    1.06 &    0.024 &    0.085 &    2.33 & 11.4 \\ 
 4663 &     1.54 &    0.24 &    0.100 &    4.27 &    3.42 &    2.29 &    2.84 &    0.030 &    0.059 &    1.96 & 17.0 \\ 
 4682 &  1.31 &    0.17 &    0.044 &    3.33 &    3.15  &    2.39 &    5.4  &   $-$0.033 &    0.059 &    3.61 & 9.0 \\
 4834 &     1.53 &    0.05 &    0.011 &    1.79 &    0.98 &    1.08 &    2.11 &   $-$0.031 &    0.035 &  \nodata & 13.0 \\ 
 4852 &     2.06 &    0.32 &    0.153 &    6.32 &    2.94 &    2.94 &    5.09 &    0.034 &    0.088 &  \nodata & 10.8 \\ 
 4864 &  1.92 &    0.18 &    0.078 &    3.60 &    0.20 &    1.56 &    0.76 &    0.005 &   $-$0.008 &  \nodata & 22.1 \\ 
 5003 &     1.62 &    0.11 &    0.019 &    2.89 &    1.09 &    1.69 &    0.67 &    0.029 &    0.056 &  \nodata & 15.2 \\ 
 5018 &     1.19 &    0.36 &    0.165 &    5.34 &    3.78 &    2.87 &    7.98 &    0.050 &    0.103 &  \nodata & 12.0 \\ 
 5097 &     1.27 &    0.46 &    0.225 &    7.47 &    3.37 &    3.25 &    9.19 &    0.070 &    0.100 &   \nodata & 12.2 \\ 
 5217 &     2.80 &    0.07 &   $-$0.008 &    1.36 &    0.60 &    2.29 &    1.21 &    0.000 &    0.011 &  \nodata & 15.5 \\ 
 6051 &  3.61 &    0.12 &    0.042 &    3.43 &    1.94 &    1.39 &    1.35 &    0.000 &    0.658 &   \nodata & 15.6 \\
\enddata
\tablenotetext{a}{from \cite{geisler96} }
\tablenotetext{b}{Signal to noise per spectral pixel over the region 5950 to 6250~\AA.}
\end{deluxetable}

\begin{deluxetable}{lccrrr}
\tablenum{3}
\tablewidth{0pt}
\small
\tablecaption{Coefficients of Index-Index 
Fits\tablenotemark{a}\label{table_coef}}
\tablehead{\colhead{X} & 
\colhead{Y} & \colhead{Sample} & \colhead{A} & \colhead{B} & \colhead{C} \\ 
\colhead{(Index)} &  \colhead{(Index)} &  
\colhead{} &  \colhead{} }
\startdata 
Mgb & Na & All &       1.674  & $-0.660$  & 0.225  \\
Mgb & $<$Fe$>$ & All & 0.927  & 0.241  & 0.023  \\
Mgb & H$\beta$ & All & 2.010  & 0.0113  & $-0.037$ \\
Mgb & H$\alpha$ & All\tablenotemark{b} & 2.728 & $-0.286$ & 0.010 \\
H$\beta$ & H$\alpha$ & High SNR\tablenotemark{b} & 0.795 & 0.624  & $-0.003$ \\
\enddata
\tablenotetext{a}{Fits to $Y = A + Bx + Cx^2$.}
\tablenotetext{b}{Only 34 of the 47 M49 GCs in our sample have
measured H$\alpha$ indices, 7 of which are highly uncertain.}
\end{deluxetable}

\begin{deluxetable}{lccrrr}
\tablenum{4}
\tablewidth{0pt}
\small
\tablecaption{Radial Distributions of the Spectroscopic GC Samples in M49 and in M87
\label{table_radial}}
\tablehead{\colhead{Galaxy} & 
\colhead{Min $R_{GC}$} & \colhead{1st Quartile} & \colhead{Median $R_{GC}$} 
& \colhead{3rd Quartile} & \colhead{Median $R_{GC}$/$R_S$}\tablenotemark{b} \\ 
\colhead{} &  \colhead{(Arcsec)} &  \colhead{(Arcsec)} & \colhead{(Arcsec)} &
\colhead{(Arcsec)} &  \colhead{} }
\startdata 
M49 & 33 & 107 & 159 & 224 & 1.0 \\
M87\tablenotemark{a} & 24 & 120 & 214 & 313 &  1.7 \\
\enddata
\tablenotetext{a}{Data from \cite{cohen98}.}
\tablenotetext{b}{Galaxy scale lengths from \cite{cote03} for M49 and from \cite{cote01}
for M87.}
\end{deluxetable}

\begin{deluxetable}{lrrrrrr}
\tablenum{5}
\tablewidth{0pt}
\small
\tablecaption{95\% Range for Indices and Abundances in 
the GC Systems of M49, M87 and the Galaxy \label{table_range}}
\tablehead{\colhead{} & \colhead{M49} & \colhead{} & \colhead{M87} 
& \colhead{} & \colhead{Galaxy} & \colhead{NGC6553,}\\
\colhead{Index} & \colhead{Min} & \colhead{Max} & \colhead{Min} & 
\colhead{Max} & \colhead{M92} & \colhead{NGC 6528} \\ 
\colhead{} &  \colhead{(\AA\ or dex)} &  \colhead{(\AA\ or dex)} 
&  \colhead{(\AA\ or dex)}
&  \colhead{(\AA\ or dex)} & \colhead{(\AA\ or dex)} & 
\colhead{(\AA\ or dex)} }
\startdata 
Mgb & 0.67 & 7.32 & 0.39 & 5.19 & 0.57 & 3.80 \\
$<$Fe$>$ & 0.10\tablenotemark{a} & 3.70 & 0.28 & 3.40 & 0.45 &  2.69 \\
NaD & 0.46 & 7.98 & 0.40 & 6.35 & 0.67 & 4.16\tablenotemark{b} \\
H$\beta$ & 0.14 & 2.87 & 0.67 & 2.98 & 2.53 & 1.70 \\
H$\alpha$ & 0.81 & 3.25 & 0.97 & 3.39 & 2.35 & 1.29 \\
$[$Z/H$]$ & $-$1.99 & 0.90\tablenotemark{c} & ... & ... & $-2.76$ & 
$-$0.03\tablenotemark{d} \\
$[$Fe/H$]$(Zinn) & $-1.90$ & 0.41 & $-2.25$ & 0.19 & $-$2.24 & $-$0.08 \\
\enddata
\tablenotetext{a}{This would increase to $\sim$0.5 if negative
equivalent widths for the Fe5335 index were ignored.}
\tablenotetext{b}{Na D lines may be affected by interstellar absorption.}
\tablenotetext{c}{This decreases to 0.85 dex when only
the Mgb and $<$Fe$>$ indices are used, ignoring the very strong NaD lines
found in the strongest lined of the M49 GCs.}
\tablenotetext{d}{The NaD indices have been ignored here.}
\end{deluxetable}

\begin{deluxetable}{lrrr}
\tablenum{6}
\tablewidth{0pt}
\small
\tablecaption{Metallicities for M49 GCs \label{table_abund}}
\tablehead{\colhead{ID\tablenotemark{a}} & 
\colhead{[Z/H]\tablenotemark{b}} & 
\colhead{Error\tablenotemark{c}} & 
\colhead{[Fe/H]$_Z$\tablenotemark{d}} \\
\colhead{} & \colhead{(dex)} & \colhead{(dex)} & \colhead{(dex)} }
\startdata 
 1475 &    $-$0.94 &     0.39 & $-$1.12 \\
 1508 &     0.32 &     0.46 & 0.11 \\
 1650 &     0.36 &     0.20 & $-$0.04 \\
 1731 &     0.10 &     0.29 & $-$0.38 \\
 1798 &     0.56 &     0.15 & 0.23 \\
 1846 &     0.54 &     0.27 & 0.22 \\
 1889 &    $-$1.99 &     0.43 & $-$1.90 \\
 1892 &    $-$1.36 &     1.36 & $-$1.66 \\
 1905 &    $-$0.56 &     0.32 & $-$0.63 \\
 2013 &    $-$1.93 &     0.78 & $-$1.22 \\
 2031 &    $-$1.04 &     0.44 & $-$0.98 \\
 2045 &    $-$0.26 &     0.32 & $-$0.48 \\
 2060 &    $-$1.88 &     0.49 & $-$1.43  \\
 2178 &    $-$1.30 &     0.82 & $-$1.22 \\
 2188 &    $-$0.68 &     0.51 & $-$0.86 \\
 2306 &    $-$0.60 &     0.29 & $-$0.68 \\
 2406 &     0.50 &     0.19 & 0.19 \\
 2421 &    $-$1.16 &     0.70 & $-$1.10 \\
 2502 &    $-$0.06 &     0.22 & $-$0.24 \\
 2528 &    $-$0.98 &     0.31 & $-$1.03 \\
 2543 &    $-$1.06 &     0.54 & $-$1.12 \\
 2569 &     0.38 &     0.12 & 0.05 \\
 2813 &     0.36 &     0.31 & 0.00 \\
 3150 &    $-$0.06 &     0.22 & $-$0.29 \\
 3603 &     0.24 &     0.29 & $-$0.14 \\
 3788 &     0.24 &     0.49 & $-$0.25 \\
 3900 &     0.80\tablenotemark{e} &     0.26 & 0.35 \\
 4017 &    $-$0.98 &     0.36 & $-$0.92 \\
 4062 &     0.90 &     0.33 & 0.41 \\
 4144 &    $-$1.10 &     0.36 & $-$1.32 \\
 4168 &    $-$0.56 &     0.29 & $-$0.71 \\
 4217 &    $-$0.04 &     0.26 & $-$0.27 \\
 4296 &    $-$1.14 &     0.77 & $-$1.28  \\
 4351 &    $-$2.06 &     0.99 & $-$2.05 \\
 4401 &     0.34 &     0.20 & 0.05 \\
 4513 &     0.06 &     0.15 & $-$0.24 \\
 4541 &    $-$0.70 &     0.58 & $-$0.84 \\
 4663 &    $-$0.04 &     0.17 & $-$0.29 \\
 4682 &     0.05 &     0.29 & $-$0.04 \\
 4834 &    $-$0.94 &     0.66 & $-$1.00 \\
 4852 &     0.56 &     0.20 & 0.23 \\
 4864 &    $-$0.22 &     0.14 & $-$0.69 \\
 5003 &    $-$0.52 &     0.44 & $-$0.83 \\
 5018 &     0.60\tablenotemark{f} &     0.25 & 0.32 \\
 5097 &     0.98\tablenotemark{g} &     0.30 & 0.49 \\
 5217 &    $-$1.50 &     0.51 & $-$1.25 \\
 6051 &    $-$0.26 &     0.31 & $-$0.62 \\
\enddata
\tablenotetext{a}{from \cite{geisler96} }
\tablenotetext{b}{This is the total [Z/H] obtained using the
models of \cite{thomas02} as described in the text.}
\tablenotetext{c}{The 1$\sigma$ rms error in [Z/H].}
\tablenotetext{d}{[Fe/H] on the scale of \cite{zinn84} with the
\cite{worthey94b} models
obtained following the procedures of \cite{cohen98}.}
\tablenotetext{e}{This decreases to 0.53 dex when only
the Mgb and $<$Fe$>$ indices are used, ignoring the very strong NaD lines.}
\tablenotetext{f}{This decreases to 0.45 dex when only
the Mgb and $<$Fe$>$ indices are used, ignoring the very strong NaD lines.}
\tablenotetext{g}{This decreases to 0.85 dex when only
the Mgb and $<$Fe$>$ indices are used, ignoring the very strong NaD lines.}
\end{deluxetable}

\begin{deluxetable}{llcccrr}
\tablenum{7}
\tablewidth{0pt}
\small
\tablecaption{Metallicity of Peaks in GC Systems \label{table_peaks}}
\tablehead{\colhead{Galaxy} & \colhead{M$_V$} &
\colhead{Low Peak\tablenotemark{a}} & 
\colhead{High Peak} 
& \colhead{Mean} & \colhead{Sample Size} & \colhead{Ref\tablenotemark{b}}  \\ 
\colhead{} &   \colhead{(mag)} & 
\colhead{(dex)} &  \colhead{(dex} &  \colhead{(dex)}
& \colhead{} & \colhead{} }
\startdata 
M49 (spec) & $-22.63$\tablenotemark{c} & $-1.1$ & $-0.1$ & $-0.6$  & 47 & 1 \\
M49 (Wash. phot) & \nodata & $-1.3$ & $-0.1$ & $-0.90$ &  1800 & 2 \\
M49 (HST phot) & \nodata & $-1.42$ & $-0.30$ & ... & 532 & 3 \\
M87 (spec) & $-22.41$\tablenotemark{c}  & $-$1.3 & 0.0:\tablenotemark{d} & $-$0.95  & 150 & 4 \\
M87 (Wash. phot) & \nodata & $-$1.3 & $-0.1$ & $-0.86$ & 407 & 5 \\
M87 (HST phot) & \nodata & $-1.40$ & $-0.30$ & $-0.84$ & 709 & 3,6 \\
NGC~1399 (spec) & $-21.3$\tablenotemark{c} & \nodata & \nodata & $-0.80$ & 18 & 7 \\
NGC~1399 (Wash. phot) & \nodata & $-1.29$ & $-0.18$ & ... & 2864 & 8 \\
NGC~1399 (HST phot) & \nodata & $-1.38$ & $-0.39$ & ... & 408 & 3 \\
M31 (spec) & $-21.2$\tablenotemark{c} & $-$1.44 & $-$0.50 & $-1.21$ & 301 & 9 \\
Galaxy &  $-20.61$\tablenotemark{e} & $-1.59$ & $-0.56$ & $-1.27$ & 147 & 10 \\
\enddata
\tablenotetext{a}{[Fe/H] on the \cite{zinn84} metallicity scale is given.}
\tablenotetext{b}{1.  This paper 2.  \cite{geisler96}, 3. \cite{larsen01} ,
4. \cite{cohen98},  5. \cite{lee93},
6. \cite{whitmore95}, 7. \cite{kissler98}, 8. \cite{dirsch03},
9. \cite{perrett02}, 10. \cite{harris96}.
The conversion of V-I color to [Fe/H] of \cite{barmby00} is used 
for the HST data.}
\tablenotetext{c}{Apparent mag from \cite{devauc91}, assuming distance of
0.77 Mpc for M31 and 16 Mpc for Virgo.  For NGC~1399, we adopt
their $v_r$ with $H_0$ of 71 km s$^{-1}$ Mpc$^{-1}$; see \cite{freedman01}.}
\tablenotetext{d}{Uncertain as the spectroscopic 
sample in M87 is dominated by GCs from the metal poor peak.}
\tablenotetext{e}{\cite{devauc78}}
\end{deluxetable}

\begin{deluxetable}{crrr}
\tablenum{8}
\tablewidth{0pt}
\small
\tablecaption{Ages of the M49 GCs Using the 
\cite{worthey94b} Models \label{table_age}}
\tablehead{\colhead{$[$Fe/H$]_Z$ Range} & \colhead{No.} & 
\colhead{Median Age} & \colhead{$\sigma$} \\
\colhead{(dex)} &  \colhead{} &  \colhead{(Gyr)} & \colhead{(Gyr)}  }
\startdata 
$ > 0.1$  & 10 & 16 & 2.8 \\
$-0.29$ to 0.05 & 10 & 13 & 4.3 \\
$-0.84$ to $-0.29$ & 10 & 11  & 3.7 \\
$-1.22$ to $-0.85$ & 10 & 13 &  2.4 \\
$< -1.26$ & 7 & 14 & 1.6 \\
$< -0.6$ & 23 & 13.9 & 4.1 \\
$> -0.6$ & 24 & 12.8 & 2.6 \\
\enddata
\end{deluxetable}

\begin{deluxetable}{crrr}
\tablenum{9}
\tablewidth{0pt}
\small
\tablecaption{Ages of the M49 GCs Using the 
\cite{thomas02} Models \label{table_tmbage}}
\tablehead{\colhead{$[$Z/H$]$ Range} & \colhead{No.} & 
\colhead{Median Age} & \colhead{$\sigma$} \\
\colhead{(dex)} &  \colhead{} &  \colhead{(Gyr)} & \colhead{(Gyr)}  }
\startdata 
$ > +0.7$  & 8 & 5 & $-2$,+6 \\
+0.3 to +0.7 & 8 & 8 & 3 \\
$-0.5$ to +0.3 & 8 & 10  & 3 \\
$-0.85$ to $-0.5$ & 8 & 15 &  3 \\
$< -0.85$ & 8 & $>15$ & $\sim2$ \\
\enddata
\end{deluxetable}

\clearpage

\clearpage

\clearpage

\clearpage

\clearpage

\clearpage


\clearpage

\clearpage

\clearpage

\clearpage

\clearpage

\clearpage

\clearpage

\clearpage

\clearpage

\clearpage

\clearpage

\clearpage

\end{document}